\documentclass[a4paper,12pt,DIV15]{scrartcl}
\usepackage{amssymb}
\usepackage{amsmath}
\usepackage{mathtools}
\usepackage{bbold}
\usepackage{slashed}
\usepackage{graphicx}
\usepackage{bbm}
\usepackage{subfigure}
\usepackage{multirow}
\usepackage{pifont}
\usepackage{authblk}
\usepackage{enumerate}
\usepackage{lscape}
\usepackage[square,sort&compress,numbers]{natbib}
\usepackage{xcolor}
\definecolor{dark-green}{rgb}{0.1,0.7,0.3}
\usepackage[colorlinks=true,linkcolor=red,urlcolor=blue,citecolor=green,menucolor=black,hyperfootnotes,linktocpage]{hyperref}

\newcommand\T{\rule{0pt}{2.8ex}} 
\newcommand\Tbig{\rule{0pt}{4.5ex}} 

\newcommand{\mee}{\langle m_{ee}\rangle}
\newcommand{\sumnu}{\sum m_i}
\newcommand{\mbeta}{m_\beta}

\newcommand{\dma}{\Delta m_{\rm A}^2}

\newcommand{\obb}{0\nu\beta\beta}

\newcommand{\mwl}{m_{W_L}}
\newcommand{\mwr}{m_{W_R}}
\newcommand{\mdl}{m_{\delta^{++}_L}}
\newcommand{\mdr}{m_{\delta^{++}_R}}

\newcommand{\muegam}{\mu\to e\gamma}
\newcommand{\mueee}{\mu\to 3e}
\newcommand{\mue}{\mu\to e}

\newcommand{\ba}{\begin{array}{c}}
\newcommand{\baz}{\begin{array}{cc}}
\newcommand{\bad}{\begin{array}{ccc}}
\newcommand{\bav}{\begin{array}{cccc}}
\newcommand{\baf}{\begin{array}{ccccc}}
\newcommand{\ea}{\end{array}}

\newcommand{\onbb}{neutrinoless double beta decay}

\def\be{\begin{equation}}
\def\ee{\end{equation}}
\def\gs{\mathrel{
   \rlap{\raise 0.511ex \hbox{$>$}}{\lower 0.511ex \hbox{$\sim$}}}}
\def\ls{\mathrel{
   \rlap{\raise 0.511ex \hbox{$<$}}{\lower 0.511ex \hbox{$\sim$}}}}

\newcommand{\bea}{\begin{equation} \begin{array}{c}}
\newcommand{\eea}{ \end{array} \end{equation}}

\newcommand{\csection}[1]{\section*{\centering #1}}
 
\def\slc#1{\setbox0=\hbox{$#1$}           
    \dimen0=\wd0                                 
    \setbox1=\hbox{/} \dimen1=\wd1               
    \ifdim\dimen0>\dimen1                        
       \rlap{\hbox to \dimen0{\hfil/\hfil}}      
       #1                                        
    \else                                        
       \rlap{\hbox to \dimen1{\hfil$#1$\hfil}}   
       /                                         
    \fi}

\title{ \Large \sffamily\bfseries Lepton number and flavour violation in TeV-scale left-right symmetric theories with large left-right mixing}
\author{\sf James Barry\footnote{E-mail: {\tt james.barry@mpi-hd.mpg.de}} }
\author{\sffamily~Werner Rodejohann\footnote{E-mail: {\tt werner.rodejohann@mpi-hd.mpg.de}} }
\affil{Max-Planck-Institut f\"{u}r Kernphysik, \\ Saupfercheckweg 1, 69117 Heidelberg, Germany}
\date{}
\begin{document}

\maketitle
\begin{abstract}
\noindent The various diagrams leading to neutrinoless double beta decay in the left-right symmetric model have different relative magnitudes, depending on the scale of new physics. Neutrinos acquire
mass from both type~I and/or type~II seesaw terms, making an unambiguous analysis difficult. We study the half-life for double beta decay in the case of type~II and type~I dominance, in the former
case including interference terms. If the heavy neutrinos of the type~I seesaw model are at the TeV scale, certain processes can be enhanced. In particular, there are regions of parameter space in
which the so-called $\lambda$- and $\eta$-diagrams can give sizable contributions to the half-life for the decay. We perform a detailed study of one such scenario, paying careful attention to
constraints from lepton flavour violation.
\end{abstract}

\newpage
\section{Introduction}

Neutrinoless double beta decay ($\obb$) is a lepton number violating process, which, if observed, would prove that neutrinos are Majorana particles~\cite{Schechter:1981bd}. New physics beyond the
standard model is required to make the process observable~\cite{Duerr:2011zd}, and there are several different theoretical frameworks that could provide the necessary operators (see the review in
Ref.~\cite{Rodejohann:2011mu}). One of those theories is the left-right symmetric model (LRSM) \cite{Mohapatra:1974gc,Pati:1974yy,Senjanovic:1975rk,Mohapatra:1980yp,Deshpande:1990ip}, in which parity
is restored at high energies and right-handed neutrinos are naturally included as part of an $SU(2)$ doublet of the extended gauge symmetry. In that case there are a number of new physics
contributions to $\obb$, either from right-handed neutrinos or Higgs triplets, with the rate for double beta decay linked to neutrino mass. This connection can be both indirect, through the couplings
to and/or mixing with right-handed neutrinos, as well as direct, via the standard light neutrino contribution (see Refs.~\cite{Glashow,Mohapatra:1980yp,Hirsch:1996qw} for some of the first discussions
of $\obb$ in the LRSM). 

In the simplest version of the LRSM one expects the scale of parity restoration to be rather high, i.e., around the GUT scale of $10^{15}$~GeV. Indeed, if all couplings in the scalar potential of the
theory are of order one then this conclusion follows naturally \cite{Deshpande:1990ip}. Nevertheless, there is still enough freedom in parameter space to allow one to consider TeV-scale left-right
symmetry, which leads to several distinct and observable signatures in present-day experiments probing leptonic processes.  On the other hand, the quark sector of the TeV-scale model is severely
constrained, due to the presence of flavour changing neutral currents (FCNCs) induced by the neutral components of Higgs bidoublets that are introduced to break electroweak symmetry. These affect
meson mixing, $CP$ violation in meson decay and the neutron electric dipole moment, and one needs the neutral component of the Higgs bidoublet to be heavier than about $15$~TeV~\cite{Maiezza:2010ic}
to avoid conflict with experiment. The mass of the right-handed $W$-boson ($W_R$) can however still be around $3$~TeV, and current LHC data is already beginning to probe $W_R$ masses of this order
\cite{ATLAS:2012ak,CMS:2012zv}. Indeed, the latest limits from the CMS experiment are roughly $\mwr \gs 2.5$~TeV (see Fig.~\ref{fig:mne_mwr_CMS}). With right-handed neutrinos of similar mass or
lighter there are observable effects in $\obb$ and lepton flavour violation (LFV). The connection between double beta decay, LHC and lepton flavour violation has recently been studied by several
authors~\cite{Tello:2010am,Das:2012ii,Deppisch:2012nb,Chakrabortty:2012mh,Nemevsek:2012iq,Parida:2012sq,Awasthi:2013ff}. 

From the theoretical point of view, the LRSM provides a natural framework for both the type I \cite{Minkowski:1977sc,Yanagida:1979as,GellMann:1980vs,Glashow,Mohapatra:1979ia} and type II
\cite{Cheng:1980qt,Lazarides:1980nt,Magg:1980ut,Mohapatra:1980yp,Schechter:1980gr,Wetterich:1981bx} seesaw mechanisms, mediated by right-handed neutrinos and Higgs triplets, respectively. In this way
the smallness of neutrino mass is connected to the restoration of parity at high energies, and the $\obb$ process can proceed via the same mediators that lead to neutrino mass. It is however rather
difficult to pin down the mechanism by which the process occurs. A simplified case that has already been studied in the literature is that of type II seesaw dominance for $m_\nu$ \cite{Tello:2010am},
which restricts the number of parameters by making the right- and left-handed Majorana mass terms proportional to each other. We perform a detailed investigation of this case including LFV
constraints explicitly in the calculation of the $\obb$ half-life, and show that there are indeed places in parameter space where the triplet contribution can be significant and can interfere
with the other contributions.

The case of type I seesaw dominance is more complicated: there are some contributions to $\obb$ that involve the left- and right-handed sectors individually as well as others that involve both
sectors, through ``left-right mixing''. A simplified version was studied in Ref.~\cite{Chakrabortty:2012mh}, and a useful formula relating the various mass matrices of the theory was presented in
Ref.~\cite{Nemevsek:2012iq}, for the case of symmetric Dirac coupling. Since the left-right mixing is always a ratio of the Dirac and Majorana mass scales, $\obb$ processes involving left-right mixing
can be enhanced for specific Dirac mass matrices. This enhancement~\cite{Buchmuller:1991tu,Kersten:2007vk} is also required for collider signatures of the TeV-scale type I seesaw mechanism with
left-handed currents (see the review in Ref.~\cite{Chen:2011de}), and there have been several studies of related
phenomenology~\cite{Ibarra:2010xw,Ibarra:2011xn,Dinh:2012bp,LopezPavon:2012zg}\footnote{In the LRSM one can produce right-handed neutrinos at the LHC via right-handed currents~\cite{Keung:1983uu}, as
will be discussed in Section~\ref{subsect:collider}.}. In the LRSM case both the so-called $\lambda$- and $\eta$-diagrams could give 
large contributions, although the latter is further suppressed by the mixing between left- and right-handed gauge bosons. This idea has also been discussed in the context of the inverse process $e^-
e^- \rightarrow W_L^-W_R^-$ \cite{Barry:2012ga}, was further emphasized in extended seesaw versions of the LRSM~\cite{Parida:2012sq,Awasthi:2013ff} and a recent analysis of mixed diagrams at the
LHC can be found in Ref.~\cite{Chen:2013foz}. We perform a thorough analysis of the type I seesaw scenario, paying attention to the correct nuclear matrix elements for the different diagrams as well
as the often severe constraints from lepton flavour violating phenomena.

The paper is outlined as follows: in Section~\ref{sec:LR} we briefly summarize the theoretical details of the left-right symmetric model (the reader familiar with the LRSM may skip this
section), and in Section~\ref{sec:lr_0vbb_lfv} we provide a detailed discussion of the $\obb$ and LFV processes in the model. Section~\ref{sec:amp_calcs} is a quantitative analysis of the various
$\obb$ amplitudes in the limit of type~I or type~II seesaw dominance; we summarize and conclude in Section~\ref{sec:concl}. A brief comment on the correlation between $\obb$ half-lives is given
in Appendix~\ref{sect:halflife_corrs}. Details of decay widths and loop functions for LFV processes can be found in Appendix~\ref{sect:lfv_lrsm}, which the reader may skip as well; an explicit
numerical example demonstrating large left-right mixing is given in Appendix~\ref{sect:explicit_example}.

\section{The left-right symmetric model} \label{sec:LR}

In the left-right symmetric model, the Standard Model is extended to include the gauge group $SU(2)_R$ (with gauge coupling $g_R \neq g_L$), and right-handed fermions are grouped into doublets under
this group. Thus we have the following fermion
particle content under $SU(2)_L \times SU(2)_R \times U(1)_{B-L}$: 
\begin{eqnarray}
 L'_{Li} &= \begin{pmatrix}\nu'_L \\ \ell'_L\end{pmatrix}_i \, \sim
(\mathbf{2},\mathbf{1},\mathbf{-1})\, , & L'_{Ri} =
\begin{pmatrix}\nu'_R \\ \ell'_R\end{pmatrix}_i \sim
(\mathbf{1},\mathbf{2},\mathbf{-1}) \, , \\[1mm]
  Q'_{Li} &= \begin{pmatrix}u'_R \\ d'_R\end{pmatrix}_i \, \sim
(\mathbf{2},\mathbf{1},\mathbf{\frac{1}{3}})\, , & Q'_{Ri} =
\begin{pmatrix}u'_R \\ d'_R\end{pmatrix}_i \sim
(\mathbf{1},\mathbf{2},\mathbf{\tfrac{1}{3}})\, , 
\end{eqnarray}
with the electric charge given by $Q=T_L^3+T_R^3+\frac{B-L}{2}$ and
$i=1,2,3$. The subscripts $L$ and $R$ are associated with the
projection $P_{L,R} = \frac 12 (1 \mp \gamma_5)$. 
In order to break the gauge symmetry and allow Majorana
mass terms for neutrinos one introduces the Higgs triplets 
\begin{equation}
 \Delta_{L,R} \equiv \begin{pmatrix} \delta_{L,R}^+/\sqrt{2} & \delta_{L,R}^{++} \\ \delta_{L,R}^0 & -\delta_{L,R}^+/\sqrt{2} \end{pmatrix},
\end{equation}
with $\Delta_L \sim (\mathbf{3},\mathbf{1},\mathbf{2})$ and $\Delta_R
\sim (\mathbf{1},\mathbf{3},\mathbf{2})$; the electroweak symmetry is
broken by the bi-doublet scalar 
\begin{equation}
 \phi \equiv \begin{pmatrix} \phi_1^0 & \phi_2^+ \\ \phi_1^- & \phi_2^0 \end{pmatrix} \sim (\mathbf{2},\mathbf{2},\mathbf{0})\, .
\end{equation}
The relevant Lagrangian in the lepton sector is
\begin{align}
 {\cal L}^{\ell}_{Y} = &-\overline{L}'_{L}(f\phi +
\tilde{f}\tilde{\phi})L'_{R} -\overline{L}'^c_{L}i\sigma_2\Delta_L h_L L'_{L} -
\overline{L}'^c_{R}i\sigma_2\Delta_R h_R L'_{R} + {\rm h.c.}, \label{eq:lag_full_lep}
\end{align}
where $\tilde{\phi} \equiv \sigma_2\phi^*\sigma_2$; $f, g$ and $h_{L,R}$ are
matrices of Yukawa couplings and charge conjugation is defined as 
\begin{equation}
 (\psi_{L,R})^c \equiv {\cal C}\overline{\psi}_{L,R}^T=(\psi^c)_{R,L}\,, \quad {\cal C} \equiv i\gamma_2\gamma_0\, . 
\end{equation}
If one assumes a discrete LR symmetry in addition to the additional gauge symmetry, the gauge couplings become equal ($g_L=g_R=g$) and one obtains relations between the Yukawa coupling matrices in the
model. With a discrete parity symmetry ($L_L \leftrightarrow L_R$, $\phi \leftrightarrow \phi^\dagger$, $\Delta_L \leftrightarrow \Delta^*_R$) it follows that $h_L=h_R^*$, $f=f^\dagger$,
$\tilde{f}=\tilde{f}^\dagger$; with a charge conjugation symmetry ($L_L \leftrightarrow (L_R)^c$, $\phi \leftrightarrow \phi^T$, $\Delta_L \leftrightarrow \Delta_R$) we have $h\equiv h_L=h_R$,
$f=f^T$, $\tilde{f}=\tilde{f}^T$. Applying these symmetries simplifies various expressions in the model, as will be discussed later.

Making use of the gauge symmetry to eliminate complex phases, the most
general vacuum is 
\begin{gather} 
 \langle \phi\rangle = \begin{pmatrix} \kappa_1/\sqrt{2} & 0 \\ 0 &
\kappa_2e^{i\alpha}/\sqrt{2} \end{pmatrix}, \quad \langle \Delta_{L}
\rangle = \begin{pmatrix} 0 & 0 \\ v_{L}e^{i\theta_L}/\sqrt{2} & 0
\end{pmatrix}, \quad \langle \Delta_{R} \rangle = \begin{pmatrix} 0 &
0 \\ v_{R}/\sqrt{2} & 0 \end{pmatrix}. 
\end{gather}
After spontaneous symmetry breaking, the mass term for the charged
leptons is 
\begin{equation}
 {\cal L}^{\ell}_{\rm mass} = -\overline{\ell}'_L M_\ell \ell'_R + {\rm h.c.},
\end{equation}
where the mass matrix
\begin{equation}
 M_\ell = \frac{1}{\sqrt{2}}(\kappa_2e^{i\alpha}f + \kappa_1\tilde{f})
\end{equation}
can be diagonalized by the bi-unitary transformation
\begin{equation}
 \ell'_{L,R} \equiv V_{L,R}^{\ell}\ell_{L,R}\, , \quad V_L^{\ell\dagger}M_\ell V_R^{\ell} = {\rm diag}(m_e,m_\mu,m_\tau)\, .
\label{eq:v_ell_def}
\end{equation}
With a discrete parity (charge conjugation) symmetry, $M_\ell$ becomes hermitian (symmetric), so that the condition $V_L^\ell=V_R^\ell$ ($V_L^\ell={V_R^\ell}^*$) holds. In the neutrino sector we have
a type~I~+~II seesaw scenario,
\begin{align}
 {\cal L}^\nu_{\rm mass} = -\tfrac{1}{2}\overline{n'_L} M_\nu n'^c_L + {\rm h.c.}
 = -\tfrac{1}{2}\begin{pmatrix} \overline{\nu'_L} \
\overline{{\nu'_R}^c}\end{pmatrix} \begin{pmatrix} M_L & M_D \\ M_D^T &
 M_R\end{pmatrix} \begin{pmatrix} {\nu'_L}^c \\ \nu'_R\end{pmatrix} + {\rm h.c.}, 
\label{eq:lag_nu_mass}
\end{align}
with
\begin{gather}
 M_D = \frac{1}{\sqrt{2}}(\kappa_1 f + \kappa_2e^{-i\alpha}\tilde{f})\, ,
\quad M_L = \sqrt{2}v_Le^{i\theta_L}h_L\, , \quad M_R = \sqrt{2}v_R h_R\,
. \label{eq:md_ml_mr}
\end{gather}
Again, with a parity (charge conjugation) symmetry we have $M^{}_D=M_D^\dagger$ ($M^{}_D=M^T_D$). 
In the most general case the phase $\theta_L$ cannot be set to zero, but in the type~II dominance case we will study it is simply an overall phase and has no effect on the resulting neutrino mass
matrix (in the type~I dominance case it plays no role since $v_L=0$). Due to the presence of the so-called ``VEV seesaw'' relation relating the various VEVs, one expects $x \equiv v_Lv_R/\kappa^2_+ =
{\cal O}(1)$, since $x$ is a function of (order one) couplings in the scalar potential \cite{Deshpande:1990ip}. However, from a purely phenomenological point of view, $x$ can take any value between 0
and $10^{14}$ \cite{Akhmedov:2006de}. Assuming that $M_L \ll M_D \ll M_R$, the light neutrino mass matrix
can be written in terms of the model parameters as 
\begin{align}
 m_\nu = M_L - M_DM_R^{-1}M_D^T
 = \sqrt{2}v_Le^{i\theta_L}h_L - \frac{\kappa_+^2}{\sqrt{2}v_R}h_Dh_R^{-1}h_D^T\, ,
\label{eq:mnu_def}
\end{align}
where
\begin{equation}
 h_D \equiv \frac{1}{\sqrt{2}}\frac{\kappa_1f +
\kappa_2e^{-i\alpha}\tilde{f}}{\kappa_+}\,, \quad  
 \kappa_+^2 \equiv |\kappa_1|^2+|\kappa_2|^2\, .
\label{eq:kappa_def}
\end{equation}
The symmetric $6\times6$ neutrino mass matrix $M_\nu$ in
Eq.~(\ref{eq:lag_nu_mass}) is diagonalized by the unitary $6\times6$
matrix \cite{Schechter:1981cv,Grimus:2000vj,Hettmansperger:2011bt} 
\begin{equation}\label{eq:Wmatrix}
 W \equiv \begin{pmatrix} V_L^{\nu} \\ V_R^{\nu} \end{pmatrix} =
\begin{pmatrix} U & S \\ T & V\end{pmatrix} \simeq \begin{pmatrix} \mathbb{1} - \frac{1}{2}RR^\dagger &
R \\ -R^\dagger & \mathbb{1} - \frac{1}{2}R^\dagger R\end{pmatrix}
\begin{pmatrix} 
V_\nu & 0 \\ 0 & V_R \end{pmatrix}\, 
\end{equation}
to $W^\dagger M_\nu W^* = {\rm diag}(m_1,m_2,m_3,M_1,M_2,M_3)$, where
the unitary matrices $V_\nu$ and $V_R$ are defined by 
\bea
 M_L - M_DM_R^{-1}M_D^T = V_\nu\,{\rm diag}(m_1,m_2,m_3) \,V_\nu^T , \\[1mm]
M_R = V_R\,{\rm diag}(M_1,M_2,M_3)V_R^T\, ,
\label{eq:mnu_MR_def}
\eea
and the matrix $R = M^{}_DM_R^{-1} + {\cal O}(M_D^3(M_R^{-1})^3)$ describes the left-right mixing. The neutrino mass eigenstates $n = n_L + n^c_L=n^c$ are defined by
\begin{align}\begin{split}
 n'_L &= \begin{pmatrix} \nu'_L \\ {\nu'_R}^c \end{pmatrix} = W n_L =
\begin{pmatrix} U & S \\ T & V \end{pmatrix}\begin{pmatrix} \nu_L \\
N^c_R \end{pmatrix}, \label{eq:nu_rot_mass} \\[1mm] 
 n'^c_L &= \begin{pmatrix} {\nu'_L}^c \\ \nu'_R \end{pmatrix} =
W^*n^c_L = \begin{pmatrix} U^* & S^* \\ T^* & V^*
\end{pmatrix}\begin{pmatrix} \nu^c_L \\ N_R \end{pmatrix} . \end{split}
\end{align}
Note that the unitarity of $W$ leads to the useful relations
\begin{equation}\label{eq:unit}
V_L^\nu V_L^{\nu\dagger} = UU^\dagger + SS^\dagger = \mathbb{1} =
V_R^\nu V_R^{\nu\dagger}=TT^\dagger+VV^\dagger\, \quad {\rm and} \quad
V_L^\nu V_R^{\nu \dagger} = UT^\dagger+SV^\dagger=0\, ,
\end{equation}
with the unitary $3\times6$ matrices $V^\nu_L = (U \ \ S)$ and $V^\nu_R = (T \ \ V)$ defined in Eq.~\eqref{eq:Wmatrix}. 

The leptonic charged current interaction in the flavour basis is
\begin{equation}
 {\cal L}^{\rm lep}_{CC} =
\frac{g}{\sqrt{2}}\left[\overline{\ell'}\gamma^\mu P_L \nu'W_{L\mu}^- + \overline{\ell'}\gamma^\mu P_R \nu' W_{R\mu}^-\right] + {\rm h.c.}, 
\end{equation}
where
\begin{equation}
 \begin{pmatrix} W_L^{\pm} \\ W_R^{\pm} \end{pmatrix} =
\begin{pmatrix} \cos\xi & \sin\xi\, e^{i\alpha} \\ -\sin\xi\,
e^{-i\alpha} & \cos\xi \end{pmatrix} \begin{pmatrix} W_1^{\pm} \\
W_2^{\pm}\end{pmatrix}
\label{eq:wlwr_mixing}
\end{equation}
characterizes the mixing between left- and right-handed gauge
bosons, with $\tan2\xi = -\frac{2\kappa_1\kappa_2}{v_R^2-v_L^2}$. With negligible mixing the gauge boson masses become
\begin{equation}
 \mwl \simeq m_{W_1} \simeq \frac{g}{2}\kappa_+\, , \quad {\rm and} \quad \mwr \simeq m_{W_2} \simeq \frac{g}{\sqrt{2}}v_R\, , \label{eq:mw1_2}
\end{equation}
and assuming that\footnote{This is justified if one assumes no cancellations in generating quark masses \cite{Zhang:2007da}.} $\kappa_2<\kappa_1$, it follows that
\begin{equation}
 \xi \simeq -\kappa_1\kappa_2/v_R^2 \simeq -2\frac{\kappa_2}{\kappa_1}\left(\frac{\mwl}{\mwr}\right)^2,
\label{eq:zeta_def}
\end{equation}
so that the mixing angle $\xi$ is at most\footnote{Although the experimental limit is $\xi < 10^{-2}$ \cite{Beringer:1900zz}, for $\mwr = {\cal O}({\rm TeV})$ one has $\xi \ls 10^{-3}$
\cite{Langacker:1989xa}; supernova bounds for right-handed neutrinos lighter than 1 MeV are even more stringent ($\xi < 3 \times 10^{-5}$) \cite{Langacker:1989xa,Barbieri:1988av}.} the square of the
ratio of left and right scales $(L/R)^2$. Here we assume $L \simeq 10^2$ GeV corresponds to the electroweak scale and $R \simeq$~TeV to the scale of parity restoration, $v_R$. For small $\xi$ the
charged current in the mass basis becomes 
\begin{equation} 
 {\cal L}^{\rm lep}_{CC} = \frac{g}{\sqrt{2}}\left[\overline{\ell_L}\gamma^\mu K_L n_L
(W_{1\mu}^- + \xi e^{i\alpha} W_{2\mu}^-) + \overline{\ell_R}\gamma^\mu  K_R n^c_L (-\xi e^{-i\alpha} W_{1\mu}^- + W_{2\mu}^-)\right]
+ {\rm h.c.}
\label{eq:lag_cc_lr}
\end{equation}
Here $K_L$ and $K_R$ are $3\times6$ mixing matrices
\begin{equation}
 K_{L} \equiv V_L^{\ell\dagger}V_L^{\nu}\, , \quad {\rm and} \quad
K_{R} \equiv V_R^{\ell\dagger}V_R^{\nu*}\, , 
\end{equation}
connecting the three charged lepton mass eigenstates $\ell_i$ to the six
neutrino mass eigenstates $(\nu_i,N_i)^T$, ($i=1,2,3$), with [using Eq.~(\ref{eq:unit})]
$K_L K_L^\dagger = K_RK_R^\dagger = \mathbb{1}$ and $K_LK_R^T = 0$. The standard neutrino mixing matrix is just the left half of $K_L$, i.e., $U_{\rm PMNS} = V_L^{\ell\dagger} U$.

In this model one also expects a new neutral gauge boson, $Z'$, which mixes with the standard model $Z$ boson. The mass eigenstates $Z_{1,2}$ have the masses
\begin{equation}
 m_{Z_1} \simeq \frac{g}{2\cos\theta_W}\kappa_+ \simeq \frac{m_{W_1}}{\cos\theta_W}\, , \quad {\rm and} \quad m_{Z_2} \simeq \frac{g\cos\theta_W}{\sqrt{\cos2\theta_W}}v_R\simeq
\sqrt{\frac{2\cos^2\theta_W}{\cos2\theta_W}}\,m_{W_2}\, ,
\label{eq:mz1_2}
\end{equation}
where $g = e/\sin\theta_W$ and the $U(1)$ coupling constant is $g'\equiv e/\sqrt{\cos2\theta_W}$. Again one expects the mixing to be of order $(L/R)^2$, i.e.,
\begin{equation}
\phi = -\frac{1}{2}\sin^{-1}\frac{g^2\kappa_+^2\sqrt{\cos\,2\theta_W}}{2c_W^2(m_{Z_2}^2-m_{Z_1}^2)} \simeq -\frac{m_{Z_1}^2\sqrt{\cos\,2\theta_W}}{m_{Z_2}^2-m_{Z_1}^2} \simeq -
\sqrt{\cos\,2\theta_W}\left(\frac{m_{Z_1}}{m_{Z_2}}\right)^2.
\end{equation}
Eqs.~\eqref{eq:mw1_2} and \eqref{eq:mz1_2} imply that $m_{Z_2} \simeq 1.7 m_{W_2}$. The current limits  \cite{Beringer:1900zz,delAguila:2010mx} on the neutral gauge boson parameters are $m_{Z'} >
1.162 $~TeV and $|\phi| < 1.2 \times 10^{-3}$. In addition, the current limits on the doubly charged triplet masses are~\cite{ATLAS:2012hi} $m_{\delta^{\pm\pm}_L} > 409$~GeV and $m_{\delta^{\pm\pm}_R}
> 322$~GeV. The theory predicts $m_{\delta^{\pm\pm}_{L,R}} \simeq v_R$, assuming order one coupling constants in the scalar potential.

\section{\texorpdfstring{$\obb$, lepton flavor violation and collider physics}{0nubb, lepton flavor violation and collider physics}} \label{sec:lr_0vbb_lfv}

\subsection{Neutrinoless double beta decay} 
\subsubsection{Particle physics amplitudes}

Here we summarize the various possible diagrams for $\obb$ in
left-right symmetric models (for one of the first analyses on this
topic, see Ref.~\cite{Hirsch:1996qw}). The Lagrangian in Eq.~\eqref{eq:lag_cc_lr} can be written as
\begin{align}
\begin{split}
 {\cal L}^{\rm lep}_{CC} &= \frac{g}{\sqrt{2}}\sum_{i=1}^6\left[\overline{e}\,\gamma^{\mu} (K_L)_{ei}P_L n_{i} (W_{1\mu}^- + \xi e^{i\alpha} W_{2\mu}^-) \right. \\ 
 & \quad + \left. \overline{e}\,\gamma^{\mu} (K_R)_{ei}P_R n_{i}(-\xi e^{-i\alpha} W_{1\mu}^- + W_{2\mu}^-)\right] + {\rm h.c.} \\
&= \frac{g}{\sqrt{2}}\sum_{i=1}^3 \left[\overline{e_L}\gamma^{\mu}(U_{ei}\nu_{Li}+S_{ei}N^c_{Ri})(W_{1\mu}^- + \xi e^{i\alpha} W_{2\mu}^-) \right. \\ 
& \quad + \left. \overline{e_R}\gamma^{\mu}(T^*_{ei}\nu^c_{Li}+V^*_{ei}N_{Ri}) (-\xi e^{-i\alpha} W_{1\mu}^- + W_{2\mu}^-)\right] + {\rm h.c.},
\end{split}
\end{align}
where in the second line we have assumed a basis where the charged leptons are diagonal (we will use this basis from now on, thus expressing all processes in terms of the matrices $U$, $S$, $T$ and
$V$). $\obb$ amplitudes arise from second order terms in perturbation theory: it is clear that one can combine either two left-handed currents, two right-handed currents or one left- and one
right-handed current. The relevant mixing matrix element also depends on whether light or heavy neutrinos are exchanged in the process; the matrices $U$, $V$, $S$ and $T$ are (to second order in
$M_D/M_R$)
\begin{align} \begin{split}
  U &\equiv \left[\mathbb{1} - \frac{1}{2}M^{}_DM_R^{-1}(M^{}_DM_R^{-1})^\dagger\right]
V_\nu, \quad V \equiv \left[\mathbb{1} - \frac{1}{2}(M^{}_DM_R^{-1})^\dagger M^{}_DM_R^{-1}\right]V_R,\\[1mm] S &\equiv M^{}_DM_R^{-1}V_R, \quad T \equiv -(M_D M_R^{-1})^\dagger
V^{}_\nu\, , \label{eq:mix_matrices} \end{split}
\end{align}
as defined in Eq.~\eqref{eq:Wmatrix}, showing that light neutrino mixing is no longer unitary. The additional possibility of $W_L-W_R$ mixing allows for diagrams with, for instance, two left-handed
hadronic currents but one left- and one right-handed leptonic current [see Fig.~\ref{fig:obb_LR_eta}], with the corresponding suppression factor of $\tan\xi$ [Eq.~\eqref{eq:zeta_def}].

\begin{figure}[t]
 \centering
 \subfigure[${\cal A}_\nu$]{\label{fig:obb_LR_standard}
 \includegraphics[width=0.35\textwidth]{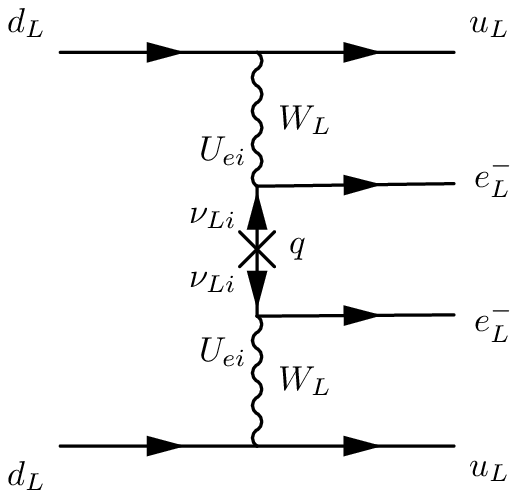}}
 \hspace{5mm}
 \subfigure[${\cal A}^R_{N_R}$]{\label{fig:obb_LR_NR}
 \includegraphics[width=0.35\textwidth]{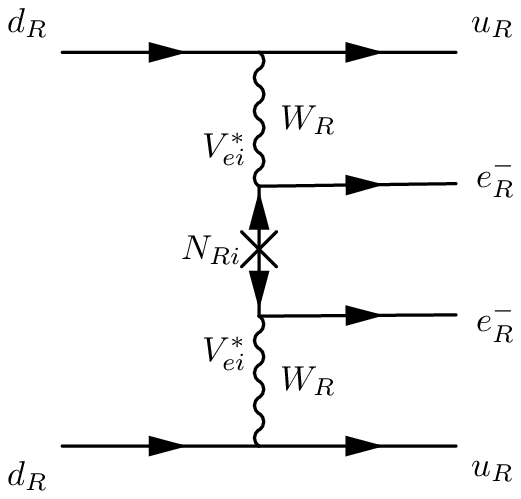}}
 \caption{Feynman diagrams of $\obb$ in the left-right symmetric model, mediated by (a) light neutrinos (the standard
mechanism ${\cal A}_\nu$) and by (b) heavy neutrinos in the presence of right-handed currents (${\cal A}^R_{N_R}$). The diagram with heavy neutrino exchange and left-handed currents (${\cal
A}^L_{N_R}$) is the same as diagram (b), with all particles left-handed and the replacement $V^*_{ei}\leftrightarrow S^{}_{ei}$. Diagrams with light neutrino exchange and right-handed currents are
negligible.} 
 \label{fig:fd_0nubb_LL_RR}
\end{figure}

Neutrinoless double beta decay processes in the LR model can be categorized in terms of their topology and the helicity of the final state electrons; the most relevant diagrams that will be discussed
in detail in what follows are shown in Figs.~\ref{fig:fd_0nubb_LL_RR}, \ref{fig:triplet_R} and \ref{fig:fd_0nubb_lambda_eta} (see Refs.~\cite{Vergados:2012xy} for a complete list).
Table~\ref{table:summary_of_mechanisms} contains a summary of the relevant amplitudes as well as limits on the particle physics parameters calculated using the recent KamLAND-Zen
limit~\cite{Gando:2012zm}\footnote{The recent GERDA limit~\cite{Agostini:2013mzu} on the half-life (see Table~\ref{table:halflifelimits}) does not improve on the limits given
here.} and the matrix elements in Table~\ref{table:matrix_elements}. Note that the chiral structure of the matrix element means that the neutrino propagator becomes
\cite{Pas:1999fc}
\begin{equation}
 P_{L,R}\frac{\slashed{q}+m_i}{q^2-m_i^2}P_{L,R} = \frac{m_i}{q^2-m_i^2×}\, \quad {\rm or} \quad P_{L,R}\frac{\slashed{q}+m_i}{q^2-m_i^2}P_{R,L} = \frac{\slashed{q}}{q^2-m_i^2×}\, ,
\end{equation}
leading to mass or momentum dependence when the leptonic vertices have the same or opposite chirality, respectively, and providing a useful way to categorize the different possible mechanisms. In
order to give a very rough estimate of the relative magnitudes we denote the masses of all particles belonging to the right-handed sector ($M_i$, $W_R$ and $\delta_R$) as $R \simeq$~TeV, and those
from the left-handed sector as $L \simeq 10^2$ GeV (corresponding to the weak scale, or the mass of the $W_L$). The matrices $T$ and $S$ describing left-right mixing can be written as $L/R$, and the
gauge boson mixing angle $\xi$ is of order $(L/R)^2$. Note that with this definition the order of magnitude of the type~I seesaw contribution is $m_\nu \simeq L^2/R$, which is far too large in the
naive case (without cancellations), but the estimates made above are still reliable. The typical scale of momentum transfer is $|q| \simeq 100$~MeV.

\minisec{Mass-dependent mechanisms}
\vspace{0.3cm}
In this case the emitted electrons have the same chirality and there are either light or heavy neutrinos exchanged, with mass denoted by $m_i$ and $M_i$. With both electrons left-handed the amplitude
is proportional to
\begin{equation}
 {\cal A}_{LL} \simeq G_F^2\left(1+2\tan\xi+\tan^2\!\xi\right)\sum_i\left(\frac{U_{ei}^2m_i}{q^2}-\frac{S_{ei}^2}{M_i}\right),
\label{eq:ll_amplitudes}
\end{equation}
whereas if both are right-handed it becomes
\begin{equation}
 {\cal A}_{RR} \simeq G_F^2\left(\frac{\mwl^4}{\mwr^4}+2\frac{\mwl^2}{\mwr^2}\tan\xi+\tan^2\!\xi\right)\sum_i\left(\frac{{T^*_{ei}}^2m_i}{q^2}-\frac{{V^*_{ei}}^2}{M_i}\right).
\label{eq:rr_amplitudes}
\end{equation}
Here we have taken into account diagrams with gauge boson mixing at one or both vertices, but the most relevant diagrams are:
\begin{itemize}
\item Fig.~\ref{fig:obb_LR_standard}, the ``standard'' diagram, with an
amplitude proportional to 
\begin{equation}
 {\cal A}_{\nu} \simeq G_F^2 \frac{\mee}{q^2}\, , 
\end{equation}
where $|q^2| \simeq (100$ MeV)$^2$ is the typical momentum exchange of the process. The particle physics parameter $|\mee|\equiv \left|\sum U_{ei}^2m_i\right|$ is called the effective mass, and the
suitably normalized dimensionless parameter that describes lepton number violation is 
\begin{equation}
|\eta_{\nu}| =  \frac{|\mee|}{m_e} = 
\frac{\left|\sum U_{ei}^2m_i\right|}{m_e} \ls 7.1 \times 10^{-7}\, ,
\end{equation}
with $U_{ei}$ the (PMNS) mixing matrix of light neutrinos and $m_i$ the light neutrino masses. Here and in what follows we give limits on the particle physics parameters $\eta_k$; they are explicitly
defined in Section~\ref{sect:NMEs}. The currently allowed~\cite{Tortola:2012te} regions of the effective mass are plotted against the lightest mass in Fig.~\ref{fig:mee_mlight}. We will translate this
plot into half-life in the following section;
\begin{figure}[t]
 \centering
 \includegraphics[angle=270,width=0.75\textwidth]{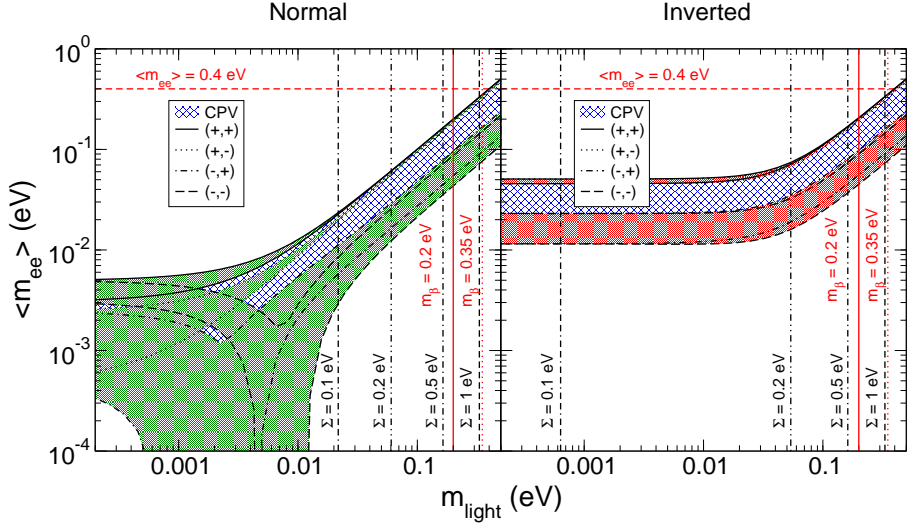}
 \caption{The effective mass $\mee$ as a function of the lightest neutrino mass in both the normal and inverted ordering, with the oscillation parameters varied in their $3\sigma$
ranges~\cite{Tortola:2012te}. CP conserving (violating) areas are indicated by black lines (blue hashes), and prospective values of $\sumnu$ and $\mbeta$ are shown.}
 \label{fig:mee_mlight}
\end{figure}
\item Fig.~\ref{fig:obb_LR_NR}, which is the analogous diagram with purely right-handed currents, mediated by right-handed
neutrinos. The amplitude is proportional to 
\begin{equation}
 {\cal A}^R_{N_R} \simeq G_F^2 \left(\frac{\mwl}{\mwr}\right)^4 \sum_i
\frac{{V^*_{ei}}^2}{M_i} \propto \frac{L^4}{R^5}\, , \label{eq:amp_RNR}
\end{equation}
where $\mwr$ ($\mwl$) is the mass of the right-handed $W_R$ (left-handed $W_L$), $M_i$ the mass of the heavy neutrinos and $V$ the right-handed analogue of the PMNS matrix $U$. The dimensionless
particle physics parameter is
\begin{equation}
\left|\eta^R_{N_R}\right| = m_p \left(\frac{\mwl}{\mwr}\right)^4 \left| \sum_i
\frac{{V^*_{ei}}^2}{M_i}\right| \ls 7.0\times 10^{-9}
\, . \label{eq:eta_RNR_def}
\end{equation}

\begin{figure}[tpb]
 \centering
 \subfigure[${\cal A}_{\delta_R}$]{\label{fig:tripletRR}
 \includegraphics[width=0.35\textwidth]{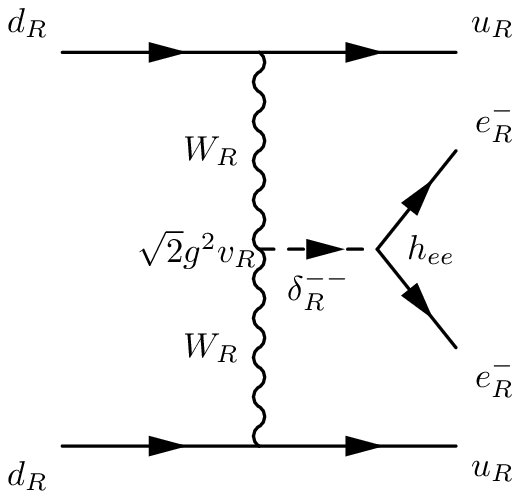}}
 \hspace{5mm}
 \subfigure[${\cal A}_{\delta_L}$]{\label{fig:tripletLL}
 \includegraphics[width=0.35\textwidth]{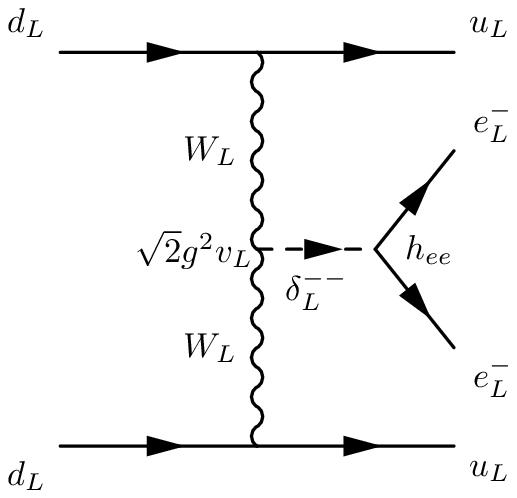}}
 \caption{Feynman diagrams of double beta decay in the left-right
symmetric model, mediated by doubly charged triplets:  (a) triplet of
$SU(2)_R$ and (b) triplet of $SU(2)_L$.} 
 \label{fig:triplet_R}
\end{figure}	

\item A diagram not shown in which heavy neutrinos are exchanged with purely left-handed currents. The amplitude is proportional to 
\begin{equation}
 {\cal A}^L_{N_R} \simeq G_F^2  \sum_i
\frac{S_{ei}^2}{M_i} \propto \frac{L^2}{R^3}\, , 
\label{eq:amp_LNR}
\end{equation}
with $S\simeq L/R$ describing the mixing of the heavy neutrinos with left-handed
currents. The limit is 
\begin{equation}
\left|\eta^L_{N_R}\right| = m_p  \left| \sum_i\frac{S_{ei}^2}{M_i} \right| 
\ls 7.0\times 10^{-9} \, . \label{eq:etalnr_lim}
\end{equation}
Note that the sum in Eq.~\eqref{eq:amp_LNR} can be written as
\begin{equation}
 \sum_i\frac{S_{ei}^2}{M_i} = \left[M_DM_R^{-1}{M_R^{-1}}^*M_R^{-1}M_D^T\right]_{ee}\, ,
\label{eq:md2mr3}
\end{equation}
which vanishes for negligible Dirac Yukawa couplings. It is also possible to have light neutrino exchange with right-handed currents [the term proportional to $T$ in Eq.~\eqref{eq:rr_amplitudes}], but
this diagram is highly suppressed.
\end{itemize}

\minisec{Triplet exchange mechanisms}
\vspace{0.3cm}
\begin{itemize}
\item Fig.~\ref{fig:tripletRR} is a diagram with different topology, mediated by the triplet of $SU(2)_R$. The amplitude is given by  
\begin{equation}
 {\cal A}_{\delta_R} \simeq G_F^2\left(\frac{\mwl}{\mwr}\right)^4\sum_i\frac{V^2_{ei}M_i}{m_{\delta^{--}_R}^2} \propto \frac{L^4}{R^5}\, ,
\label{eq:amp_triplet_R}
\end{equation}
and the dimensionless particle physics parameter is
\begin{equation}
\left|\eta_{\delta_R}\right|  = \frac{\left|\sum_i V^2_{ei}M_i\right|}{m_{\delta^{--}_R}^2\mwr^4}\frac{m_p}{G_F^2} \ls 7.0\times 10^{-9}\, .
\end{equation}
Here we have used the fact that the term $\sqrt{2}v_R h_{ee}$ is nothing but the $ee$ element of the right-handed Majorana neutrino mass matrix $M_R$ diagonalized by $V$ [cf.~Eq.~\eqref{eq:md_ml_mr}],
with $v_R$ the VEV of the triplet $\delta_R$ and $h_{ee}$ the coupling of the triplet with right-handed
electrons, so that this diagram still indirectly depends on the heavy neutrino mass;  
\begin{figure}[t]
 \centering
 \subfigure[${\cal A}_\lambda$]{\label{fig:obb_LR_lambda}
 \includegraphics[width=0.35\textwidth]{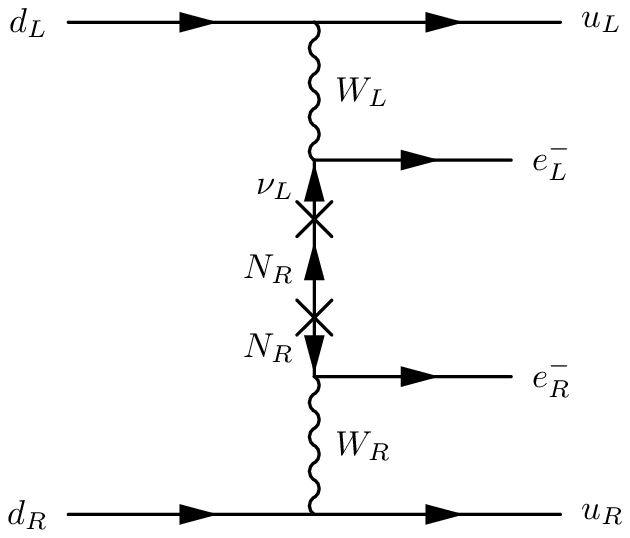}}
 \hspace{5mm}
 \subfigure[${\cal A}_\eta$]{\label{fig:obb_LR_eta}
 \includegraphics[width=0.35\textwidth]{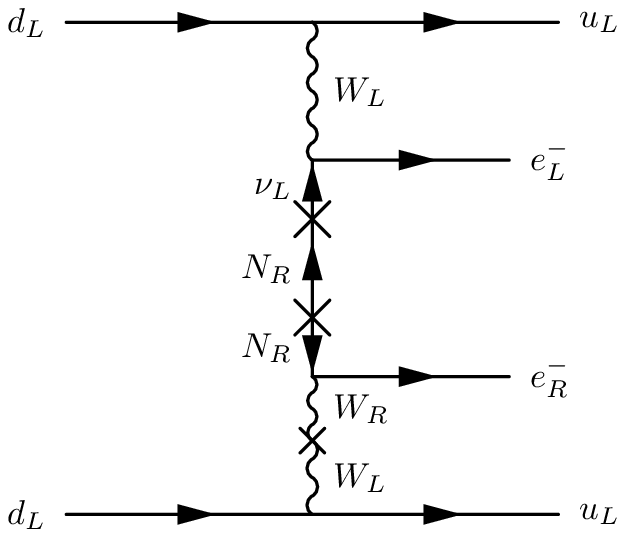}}
 \caption{Feynman diagrams of double beta decay in the left-right
symmetric model with final state electrons of different helicity: (a)
the $\lambda$-mechanism and (b) the $\eta$-mechanism due to gauge
boson mixing.} 
 \label{fig:fd_0nubb_lambda_eta}
\end{figure}
\item  Fig.~\ref{fig:tripletLL} is a diagram mediated by the triplet of $SU(2)_L$, also present in the usual type~II seesaw model (without left-right symmetry). The amplitude is given by 
\begin{equation}
 {\cal A}_{\delta_L} \simeq G_F^2 \frac{h_{ee} v_L}{m_{\delta^{--}_L}^2}\, ,
\label{eq:trip_L}
\end{equation}
which is suppressed with respect to the standard light neutrino exchange by at least a factor $q^2/m_{\delta^{--}_L}^2$.
\end{itemize}

\minisec{Momentum dependent mechanisms}
\vspace{0.3cm}
In this case the emitted electrons have opposite helicity, and the amplitude is proportional to
\begin{equation}
 {\cal A}_{LR} \simeq G_F^2 \left(\frac{\mwl^2}{\mwr^2}+\tan\xi+\frac{\mwl^2}{\mwr^2}\tan\xi+\tan^2\!\xi\right)\sum_i\left(U^{}_{ei}T^{*}_{ei}\frac{1}{q} -S^{}_{ei}V^*_{ei} \frac{q}{M_i^2}\right);
\label{eq:lr_amplitudes}
\end{equation}
the most important diagrams are those involving light neutrinos and two powers of the left-right mixing in the prefactor, i.e.,

\begin{itemize}
\item The so-called $\lambda$-diagram in Fig.~\ref{fig:obb_LR_lambda}, with an amplitude
\begin{equation}
 {\cal A}_\lambda \simeq G_F^2 \left(\frac{\mwl}{\mwr}\right)^2 \sum_i U_{ei} T^*_{ei}\, \frac{1}{q} \propto \frac{L^3}{R^3q}\, ,
\end{equation}
and particle physics parameter
\begin{equation}
\left|\eta_{\lambda}\right| = \left(\frac{\mwl}{\mwr}\right)^2\left|\sum_i U_{ei}T^*_{ei}\right| \ls 5.7 \times 10^{-7} \,.
\label{eq:etal}
\end{equation}
Note that this is a long-range diagram with light neutrinos exchanged, with the matrix $T^*_{ei} = {\cal O}(M_D/M_R)$ quantifying the mixing of light neutrinos with right-handed currents.
\item The $\eta$-diagram in Fig.~\ref{fig:obb_LR_eta}, which also has mixed helicity and light neutrino exchange (long-range diagram). This is only possible due to $W_L-W_R$ mixing, described by the
parameter $\tan\xi$ [see Eq.~\eqref{eq:wlwr_mixing}]. The amplitude is  
\begin{equation}
 {\cal A}_\eta \simeq G_F^2\tan\xi\sum_iU_{ei}T^*_{ei}\, \frac{1}{q} \propto \frac{L^3}{R^3q}\, ,
\end{equation}
with particle physics parameter
\begin{equation}
\left|\eta_{\eta}\right| = \tan\xi\left|\sum_i U_{ei}T^*_{ei}\right| \ls 3.0 \times 10^{-9}
\,.
\label{eq:etaeta}
\end{equation}
Ref.~\cite{Vergados:2002pv} gives a detailed explanation of how a complicated cancellation of different nuclear physics amplitudes leads to a limit on the $\eta$-diagram that is much stronger than the
one on the $\lambda$-diagram. The heavy neutrino contributions to both the $\lambda$- and $\eta$ diagrams are further suppressed, being proportional to $\sum_i S^{}_{ei}V^*_{ei}q/M^2_i$ [see
Eq.~\eqref{eq:lr_amplitudes}]. Using the mixing matrices in Eq.~\eqref{eq:mix_matrices}, the relevant sums become
\begin{align}
 \sum_i U^{}_{ei}T^*_{ei} &= \left[\left(\mathbb{1} - \frac{1}{2}M^{}_DM_R^{-1}(M^{}_DM_R^{-1})^\dagger\right)
V^{}_\nu\left(-(M_R^{-1})^TM_D^T V^*_\nu\right)^T\right]_{ee} \simeq -\left[M^{}_DM_R^{-1}\right]_{ee}, \notag \\
 \sum_i S^{}_{ei}V^*_{ei} &= \left[M^{}_DM_R^{-1}V_R\left(\left(\mathbb{1} - \frac{1}{2}(M^{}_DM_R^{-1})^T (M^{}_DM_R^{-1})^*\right)V_R^*\right)^T\right]_{ee} \simeq \left[M^{}_DM_R^{-1}\right]_{ee},
\end{align}
where we have omitted third order terms. This again shows that the left-right mixing is a ratio of two scales, $M_D$ and $M_R$.
\end{itemize}

Using our rough estimates (in terms of $L\simeq10^2$~GeV and $R\simeq$~TeV) of the scale of each diagram we can now make a naive guess at their expected relative magnitudes. Since the mixed $\lambda$-
and $\eta$-diagrams in Fig.~\ref{fig:fd_0nubb_lambda_eta} are of order $(L/R)^3 /q$ and the purely right-handed short-range diagrams in Figs.~\ref{fig:obb_LR_NR} (heavy neutrino exchange and
right-handed 
currents) and \ref{fig:tripletRR} ($SU(2)_R$ triplet exchange and right-handed currents) are of order $L^4 /R^5$, we expect the mixed diagrams to dominate by a factor  $R^2/(L q) \sim 10^5$. In the
same sense, the amplitudes  of the mixed diagrams are also larger than the one for heavy neutrino exchange with left-handed currents, proportional to $L^2/R^3$. However, these simple estimates
are rather optimistic since the smallness of neutrino masses means that the left-right mixing should be much smaller than $L/R \simeq 0.1$. In the absence of cancellations the mixing is
bounded as
\begin{equation}
 |S_{\alpha i}| \simeq |T^T_{\alpha i}| \simeq \sqrt{\frac{m_\nu}{M_i}} \ls 10^{-7}\left(\frac{\rm TeV}{M_i}\right)^{1/2}, \quad (\alpha=e,\mu,\tau)\, ,
\label{eq:lr_mixing_nocancel_bound}
\end{equation}
so that it is obvious that the light neutrino mass from type~I seesaw, $m_\nu \simeq M_D^2/M_R \simeq L^2/R$ cannot be small enough without special matrix structures. The crucial point is that the
left-right mixing $M_D/M_R \simeq L/R$ can still be large in some cases, which means that mixed diagrams should be examined more thoroughly, as has been done in the context of inverse
neutrinoless double beta decay \cite{Barry:2012ga} and inverse/extended seesaw \cite{Parida:2012sq,Awasthi:2013ff}. Note that the limits
on the difference of the diagonal elements of the product $\epsilon_\alpha \equiv [SS^\dagger]_{\alpha\alpha} \simeq [T^\dagger T]_{\alpha\alpha}$ from lepton universality \cite{Loinaz:2004qc} are
\begin{equation}
 \epsilon_{e} - \epsilon_\mu \ls 0.0022\,, \quad \epsilon_\mu - \epsilon_\tau \ls 0.0017\,, \quad \epsilon_e - \epsilon_\tau \ls 0.0039\, ,
\end{equation}
which give a rather weak bound on the left-right mixing.

The reliability of the rough approximations in terms of $L$ and $R$ can be tested by normalising the amplitudes to the standard contribution, using known bounds on the left-right mixing. We use the
bound in Eq.~\eqref{eq:lr_mixing_nocancel_bound} in the estimates that follow, along with the light neutrino mass scale $m_\nu \simeq 0.05$~eV and momentum exchange $|q| \simeq 100$~MeV. It turns out
that the mixed helicity diagrams ${\cal A}_\lambda$ and ${\cal A}_\eta$ can still compete with the standard light neutrino diagram, even with the stringent limit on $T$ in
Eq.~\eqref{eq:lr_mixing_nocancel_bound} that connects the left-right mixing to light neutrino mass. For heavy neutrino exchange with right-handed currents [Fig.~\ref{fig:obb_LR_NR}] we have
\begin{equation}
 \frac{{\cal A}_{N_R}^R}{{\cal A}_\nu} \simeq \left(\frac{\mwl}{\mwr}\right)^4 \sum_i\frac{{V^*_{ei}}^2}{M_i} \frac{q^2}{m_\nu} \simeq 8.36 \left(\frac{\rm TeV}{\mwr}\right)^4\left(\frac{\rm
TeV}{M_i}\right),
\end{equation}
whereas for heavy neutrino exchange with left-handed currents [Eq.~\eqref{eq:amp_LNR}] the ratio is
\begin{equation}
 \frac{{\cal A}_{N_R}^L}{{\cal A}_\nu} \simeq \sum_i\frac{S_{ei}^2}{M_i}\frac{q^2}{m_\nu} \ls \frac{q^2}{M_i^2} \simeq 10^{-8} \left(\frac{\rm TeV}{M_R}\right)^2.
\end{equation}
One sees immediately that this process requires cancellations to be enhanced\footnote{This case was also studied in Ref.~\cite{Ibarra:2010xw,Mitra:2011qr,Dinh:2012bp,LopezPavon:2012zg}.}. However, for
the $\lambda$- and $\eta$-diagrams [Fig.~\ref{fig:fd_0nubb_lambda_eta}] we have
\begin{equation}
 \frac{{\cal A}_\eta}{{\cal A}_\nu} \ls \frac{{\cal A}_\lambda}{{\cal A}_\nu} \simeq \left(\frac{\mwl}{\mwr}\right)^2 \sum_iU^{}_{ei}T^{*}_{ei} \frac{q}{m_\nu} \ls \left(\frac{\mwl}{\mwr}\right)^2
\frac{q}{\sqrt{m_\nu M_i}} \simeq 2.89 \left(\frac{\rm TeV}{\mwr}\right)^2\left(\frac{\rm TeV}{M_R}\right)^{1/2},
\end{equation}
where the first inequality comes from the upper limit $|\xi| \ls \left(\frac{\mwl}{\mwr}\right)^2$. Depending on the relative magnitude of the bidoublet VEVs $\kappa_1$ and $\kappa_2$, the amplitude
${\cal A}_\eta$ may be further suppressed [see Eq.~\eqref{eq:zeta_def}], but this could be compensated for by the fact that ${\cal M}^{0\nu}_\eta \simeq 10^2 {\cal M}^{0\nu}_\lambda$
(cf.~Table~\ref{table:matrix_elements}). The main point is that with $\mwr$ and $M_R$ around the TeV scale the amplitudes ${\cal A}^R_{N_R}$, ${\cal A}_{\lambda}$ and ${\cal A}_{\eta}$ turn out to be
quite close in magnitude, whereas the small value for ${\cal A}^L_{N_R}$ could still be enhanced by cancellations. Note that in order to arrange for this the Yukawa matrices $f$ and
$\tilde{f}$ need to have non-trivial flavour structure so that the correct light neutrino mass [see Eq.~\eqref{eq:md_ml_mr}] can be obtained, since with ${\cal O}(1)$ couplings, $M_D \propto
\kappa_i$, so that $M_D$ would be near the electroweak scale of $10^2$ GeV. Assuming that $\kappa_2 \ll \kappa_1$ (see also Ref.~\cite{Zhang:2007da}) means that $M_D \simeq  \kappa_1 f/\sqrt{2}$ and
$M_\ell \simeq \kappa_2 \tilde{f}/\sqrt{2}$, so that one has the freedom to choose $f$ without affecting the charged leptons.
\begin{table}
 \centering
 \caption{Summary of relevant mechanisms for $\obb$ in the left-right symmetric model, with limits on new physics parameters (written in bold face) in each case (see also
Ref.~\cite{Rodejohann:2011mu}).}
\label{table:summary_of_mechanisms}
\vspace{10pt}
 \begin{tabular}{ccc}
  \hline \hline \T mechanism & amplitude & current limit \\
 \hline \Tbig light neutrino exchange (${\cal A}_\nu$) & $\dfrac{G_F^2}{q^2} \boldsymbol{\left|U_{ei}^2 m_i\right|}$ & 0.36~eV \\[4mm]
 heavy neutrino exchange (${\cal A}^L_{N_R}$) & $G_F^2 \boldsymbol{\left|\dfrac{S_{ei}^2}{M_i}\right|}$ & $7.4\times 10^{-9}\ {\rm  GeV}^{-1}$ \\[4mm]
 heavy neutrino exchange (${\cal A}^R_{N_R}$) & $G_F^2\mwl^4 \boldsymbol{\left|\dfrac{{V^*_{ei}}^2}{M_i \mwr^4}\right|}$ & $1.7\times 10^{-16}\ {\rm GeV}^{-5}$ \\[4mm]
 Higgs triplet exchange (${\cal A}_{\delta_R}$) & $G_F^2\mwl^4 \boldsymbol{\left|\dfrac{V^2_{ei}M_i}{m_{\delta^{--}_R}^2\mwr^4}\right|}$ & $1.7\times 10^{-16}\ {\rm GeV}^{-5}$ \\[4mm]
 $\lambda$-mechanism (${\cal A}_\lambda$) & $G_F^2 \dfrac{\mwl^2}{q}\boldsymbol{\left|\dfrac{U_{ei} T^*_{ei}}{\mwr^2}\right|}$ & $8.8 \times 10^{-11}\ {\rm GeV}^{-2}$ \\[4mm]
 $\eta$-mechanism (${\cal A}_\eta$) & $G_F^2\dfrac{1}{q}\boldsymbol{\left|\tan\xi\sum_iU_{ei}T^*_{ei}\right|}$ & $3.0 \times 10^{-9}$ \\[4mm]
 \hline \hline
 \end{tabular}
\end{table}

\subsubsection{Nuclear matrix elements and lifetime} \label{sect:NMEs}

In order to translate the dimensionless particle physics parameters $\eta_k$ into actual lifetimes of $\obb$ processes for different isotopes one needs the relevant nuclear matrix elements and phase
space factors. There are various different methods to calculate those quantities and most previous studies have focussed on the standard light neutrino exchange mechanism; here we attempt to compile a
list of the most recently calculated matrix elements relevant to $\obb$ in the LR model, combining the calculations of various groups.

We use the QRPA calculation of the matrix elements for the mixed diagrams in Ref.~\cite{Pantis:1996py} (see also Refs.~\cite{Muto:1989cd,Suhonen:1998ck}). In their notation, the lifetime of $\obb$ can
be written as
\begin{align}
 \left[T_{1/2}^{0\nu}\right]^{-1} &= G^{0\nu}_{01}|{\cal M}^{0\nu}_{\rm GT}|^2\left\{\left|X_L\right|^2 + \left|X_R\right|^2 + \tilde{C}_2|\eta_\lambda| |X_L|\cos\psi_1+\tilde{C}_3|\eta_\eta| |X_L|
\cos\psi_2 \right. \notag \\
& \left. + \ \tilde{C}_4|\eta_\lambda|^2 + \tilde{C}_5|\eta_\eta|^2+\tilde{C}_6|\eta_\lambda||\eta_\eta|\cos(\psi_1-\psi_2)+{\rm
Re}\left[\tilde{C}_2X_R\eta_\lambda+\tilde{C}_3X_R\eta_\eta\right]\right\}, \label{eq:half-life_full}
\end{align}
where the coefficients $\tilde{C}_i$ are combinations of matrix elements and integrated kinematical factors, $G^{0\nu}_{01}$ is the usual phase space factor and $\psi_i$ are complex phases. The
parameters $X_L$ ($X_R$) include all processes in which the final state electrons are both left-handed (right-handed), i.e.
\begin{equation}
 X_L \equiv {\cal M'}^{0\nu}_\nu\eta_{\nu}+{\cal M'}^{0\nu}_N\eta^L_{N_R}+{\cal M'}^{0\nu}_N\eta_{\delta_L}\, , {\rm and} \quad X_R \equiv {\cal M'}_N^{0\nu}\eta^R_{N_R}+{\cal
M'}^{0\nu}_N\eta_{\delta_R}\, ,
\end{equation}
with $\eta_{\delta_L}$ the LNV parameter associated with Eq.~\eqref{eq:trip_L}. In Eq.~\eqref{eq:half-life_full} we have omitted the interference term $X_L X_R$, which is suppressed due to the
different electron helicities ($e^-_Le^-_L$ vs $e^-_Re^-_R$); interference terms with final states in which at least one of the electrons has the same helicity have been included. The matrix elements
${\cal M'}^{0\nu}_\nu$ and ${\cal M'}^{0\nu}_N$ include Fermi and Gamow-Teller contributions.

Ref.~\cite{Kotila:2012zza} presents an improved calculation of the phase space factor $G^{0\nu}_{01}$ for the light neutrino exchange mechanism, taking into account the finite nuclear size of the
Dirac wave function as well as electron screening effects and angular correlations. The factor is slightly lower, with the difference becoming more marked for heavier nuclei. The coefficients
$\tilde{C}_i$ ($i=2,3,4,5,6$) depend on different phase space factors~\cite{Doi:1985dx,Pantis:1996py}; here we assume those factors are reduced by the same percentage as $G^{0\nu}_{01}$. More recent
calculations \cite{Simkovic:1999re,Faessler:2011qw,Faessler:2011rv} of light and heavy neutrino matrix elements include the Gamow-Teller factor ${\cal M}^{0\nu}_{\rm GT}$ in the relevant matrix
elements ${\cal M}^{0\nu}_\nu$ and ${\cal M}^{0\nu}_N$. For consistency of notation, we make the following definitions 
\begin{align}
 \begin{split}
{\cal M}^{0\nu}_\nu \equiv {\cal M}^{0\nu}_{\rm GT}{\cal M'}^{0\nu}_\nu\, , \quad {\cal M}^{0\nu}_N \equiv {\cal M}^{0\nu}_{\rm GT}{\cal M'}^{0\nu}_N\, , \\ {\cal M}^{0\nu}_\lambda \equiv \sqrt{|{\cal
M}^{0\nu}_{\rm GT}|^2\tilde{C}_4}\, , \quad {\cal M}^{0\nu}_\eta \equiv \sqrt{|{\cal M}^{0\nu}_{\rm GT}|^2\tilde{C}_5}\, , 
\end{split}
\end{align}
which allow us to write the lifetime in Eq.~\eqref{eq:half-life_full} as
\begin{align}
 \left[T_{1/2}^{0\nu}\right]^{-1} &= G^{0\nu}_{01}\left\{|{\cal M}^{0\nu}_\nu|^2|\eta_\nu|^2 + |{\cal M}^{0\nu}_N|^2|\eta^L_{N_R}|^2 + |{\cal M}^{0\nu}_N|^2|\eta^R_{N_R}+\eta_{\delta_R}|^2  \right.
\notag \\ & \left. + |{\cal M}^{0\nu}_\lambda|^2|\eta_\lambda|^2 + |{\cal M}^{0\nu}_\eta|^2|\eta_\eta|^2 \right\} + {\rm interference\ terms}. \label{eq:half-life_simp}
\end{align}
The corresponding matrix elements are reported in Table~\ref{table:matrix_elements} and will be used in the analysis that follows. The range of values comes from the fact that different calculations
have been used. Note that we have used the new phase space numbers to calculate limits.
\begin{table}[tpb]
 \centering
 \caption{The phase-space factor $G^{0\nu}_{01}$~\cite{Pantis:1996py,Kotila:2012zza} and the matrix elements for light (${\cal M}^{0\nu}_\nu$)~\cite{Dueck:2011hu} and heavy (${\cal
M}^{0\nu}_N$)~\cite{Faessler:2011qw,Faessler:2011rv,Vergados:2012xy,Meroni:2012qf} neutrino exchange, and for the $\lambda$- and $\eta$-diagrams~\cite{Pantis:1996py,Suhonen:1998ck}, for different
isotopes, for $g_A = 1.25$ and $r_0 = 1.1$~fm, corresponding to Eq.~\eqref{eq:half-life_simp}.}
\label{table:matrix_elements}
\vspace{10pt}
 \begin{tabular}{lccccccc}
 \hline \hline
 \T \multirow{2}{*}{Isotope} & $G^{0\nu}_{01}$ $[10^{-14}\ {\rm yrs}^{-1}]$ & $G^{0\nu}_{01}$ $[10^{-14}\ {\rm yrs}^{-1}]$ & \multirow{2}{*}{${\cal M}^{0\nu}_\nu$} & \multirow{2}{*}{${\cal
M}^{0\nu}_N$} & \multirow{2}{*}{${\cal M}^{0\nu}_\lambda$} & \multirow{2}{*}{${\cal M}^{0\nu}_\eta$} \\[1mm] 
 & (old~\cite{Pantis:1996py}) & (new~\cite{Kotila:2012zza}) & & & \\[0.8mm]
\hline \T 
$^{76}$Ge & 0.793 & 0.686 &2.58--6.64 & 233--412 & 1.75--3.76 & 235--637 \\ 
$^{82}$Se & 3.53 & 2.95 &2.42--5.92 & 226--408 & 2.54--3.69 & 209--234 \\ 
$^{130}$Te & 5.54 & 4.13 &2.43--5.04 & 234--385 & 2.85--3.67 & 414--540 \\ 
$^{136}$Xe & 5.91 & 4.24 &1.57--3.85 & 164--172 & 1.96--2.49 & 370--419 \\ 
\hline \hline
 \end{tabular}
\end{table}

\begin{figure}[t]
 \centering
 \includegraphics[angle=270,width=0.8\textwidth]{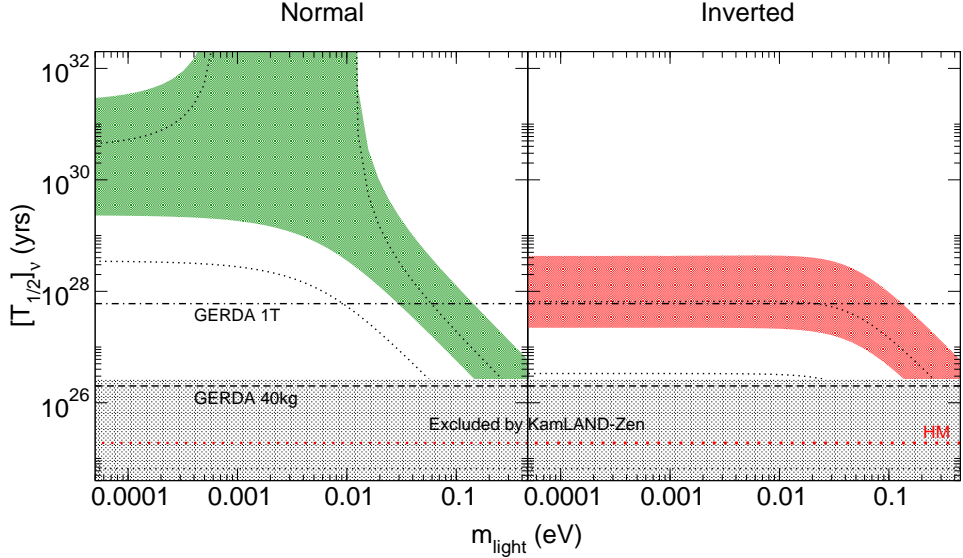}
 \caption{The standard light neutrino contribution to the $\obb$ half-life of $^{76}$Ge plotted against the lightest light neutrino mass, using $3 \sigma$ ranges of the oscillation data from
Ref.~\cite{Fogli:2012ua}. Shaded regions (dotted lines) are for the smallest (largest) NMEs from Table~\ref{table:matrix_elements}. The grey shaded region is excluded by the KamLAND-Zen
experiment, the horizontal dashed (dashed-dotted) lines show the planned sensitivities of the GERDA~\cite{Ackermann:2012xja} experiment, with 40 kg (1 ton) of isotope. The Heidelberg-Moscow
limit~\cite{KlapdorKleingrothaus:2000sn} is indicated by a horizontal (red) dotted line. The variation in the KamLAND-Zen limit due to the NMEs for $^{76}$Ge and $^{136}$Xe is shown by the dotted
black horizontal line, which is the minimum value this limit can take.}
\label{fig:standard_lifetime_Ge}
\end{figure}

In the limit of type~II seesaw dominance, the expression in Eq.~\eqref{eq:half-life_simp} will simplify considerably, whereas with type I seesaw dominance all six terms should be considered [we
neglect the contribution stemming from the left-handed triplet $\delta_L$, which is suppressed by light neutrino mass and $m_{\delta^{--}_L} = {\cal O}({\rm TeV})$]. We use the notation
$[T^{0\nu}_{1/2}]_k$ ($k = \nu,N^{(R)}_R,N^{(L)}_R,\delta_R,\lambda,\eta$) to refer to the lifetime stemming from one particular diagram. Figure~\ref{fig:standard_lifetime_Ge} illustrates the
variation of the lifetime $[T^{0\nu}_{1/2}]_\nu$ with lightest neutrino mass, for the $\obb$ of $^{76}$Ge and using both the smallest and largest matrix element (${\cal M}^{0\nu}_\nu = 2.58$);
comparison with Fig.~\ref{fig:mee_mlight} shows that the lifetime is obviously just the inverse of the effective mass, with various numerical prefactors. The variation in ${\cal M}^{0\nu}_\nu$ can
bring the minimum allowed lifetime down by roughly one order 
of magnitude.

\subsection{Charged lepton flavor violation and dipole moments} \label{subsect:lfv_lrsm}

Although small active neutrino masses ``GIM suppress'' charged lepton flavor violating processes by a factor of $(\dma/\mwl^2)^2 \ls 10^{-50}$ ($\dma$ is the atmospheric mass squared difference), the
existence of heavy right-handed neutrinos and Higgs scalars allow the LFV decays $\muegam$ and $\mueee$ as well as $\mue$ conversion in nuclei to occur at rates observable in current experiments.
Those decay rates will be proportional to similar combinations of mass and mixing parameters as the $\obb$ amplitudes, thus providing complementary constraints. Defining 
\begin{equation}
\Gamma_\nu \equiv \Gamma(\mu^-\to e^-\nu_\mu\bar{\nu}_e) \quad \rm{and} \quad \Gamma_{\rm capt} \equiv \Gamma(\mu^-+A(Z,N)\to \nu_\mu+A(Z-1,N+1)), 
\end{equation}
the relevant branching ratios
\begin{align}
 {\rm BR}_{\muegam} &\equiv \frac{\Gamma(\mu^+\to e^+\gamma)}{\Gamma_\nu}\, , \notag \\
 {\rm R}^A_{\mue} &\equiv \frac{\Gamma(\mu^-+A(N,Z)\to e^-+A(N,Z))}{\Gamma_{\rm capt}} \, , \label{eq:br_defs}\\
 {\rm BR}_{\mueee} &\equiv \frac{\Gamma(\mu^+\to e^+e^-e^+)}{\Gamma_\nu}\, , \notag
\end{align}
are constrained at 90\% C.L. to
$$
 {\rm BR}_{\muegam} < 5.7 \times 10^{-13} \ \mbox{\cite{Adam:2013mnn}}, 
\quad  {\rm R}^{\rm Au}_{\mue} < 7.0 \times 10^{-13} \ \mbox{\cite{Bertl:2006up}}
\quad {\rm and} \quad {\rm BR}_{\mueee} < 1.0 \times 10^{-12}\ \mbox{\cite{Bellgardt:1987du}}
$$
by experiment.

The amplitudes for LFV decays in the LRSM receive contributions from $(i)$ right-handed gauge bosons and Higgs triplets, suppressed by $(\mwl/\mwr)^2$; $(ii)$ left-handed gauge bosons, suppressed by
$\simeq|M^{}_DM_R^{-1}|^2$ and $(iii)$ processes with $W_L-W_R$ mixing, suppressed by $\xi M_DM_R^{-1}$. Terms proportional to $\xi^2$ are expected to be small and are neglected here. All of the
possible channels are in some way related to the right-handed neutrino mass, either directly as a virtual particle in the loop or indirectly since the couplings of the triplets to leptons are
proportional to $M_R$\footnote{The assumption of a discrete left-right symmetry means that the exchange of left-handed triplets is also related to right-handed neutrino mass, see
Eq.~\eqref{eq:manifest_LRSM}.}. 

A detailed calculation of the LFV decay widths and branching ratios in the LRSM has been performed in Ref.~\cite{Cirigliano:2004mv}, where the results have been obtained by expanding to leading order
in the ratios $M_D/M_R$ and $\kappa_+/v_R$, and thus ignore any effects of left-right mixing. The results are (see also Refs.~\cite{Leontaris:1985qc,Swartz:1989qz})
\begin{equation}
 {\rm BR}^{\rm triplet}_{\mueee} = \frac{1}{8}\left|\tilde{h}^{}_{\mu e}\tilde{h}^*_{ee}\right|^2\left(\frac{\mwl^4}{m_{\delta^{++}_L}^4}+\frac{\mwl^4}{m_{\delta^{++}_R}^4} \right), 
\label{eq:brmueee_tree}
\end{equation}
for the tree-level process $\mueee$ and
\begin{align}
 {\rm BR}_{\muegam} &\simeq 1.5 \times 10^{-7}\, |g_{\rm lfv}|^2\left(\frac{1\ {\rm TeV}}{\mwr}\right)^4, \\
 {\rm R}^{\rm Au}_{\mue} &\simeq 8 \times 10^{-8}\, |g_{\rm lfv}|^2\left(\frac{1\ {\rm TeV}}{m_{\delta^{++}_{L,R}}}\right)^4\alpha\left(\log\frac{m^2_{\delta^{++}_{L,R}}}{m_\mu^2}\right)^2,
\end{align}
for the loop-suppressed decays $\muegam$ and $\mue$ conversion, where the expressions are simplified by assuming the ``commensurate mass spectrum'' $M_i \simeq \mwr \simeq \mdl \simeq \mdr \simeq
m_{\delta^{+}_R}$. 
The parameters $\tilde{h}$ and $g_{\rm lfv}$ are defined to leading order in the ratio $M_D/M_R$ by
\begin{gather}
 \tilde{h}_{\alpha\beta} \equiv \sum_{i=1}^3 V_{\alpha i}V_{\beta i}\frac{M_i}{\mwr} 
 = \frac{\left[M_R\right]_{\alpha\beta}}{\mwr} \quad {\rm and} \quad 
 g_{\rm lfv} \equiv \sum_{i=1}^3 V^{}_{\mu i} V^*_{e i}\left(\frac{M_i}{\mwr}\right)^2 = \frac{\left[M_RM_R^*\right]_{\mu e}}{\mwr^2}\, , 
\label{eq:lfv_params}
\end{gather}
assuming manifest left-right symmetry (i.e., a discrete parity symmetry, see the Appendix). If one assumes that logarithmic terms [see Eq.~\eqref{eq:fgamm_form}] from doubly charged Higgs diagrams
dominate and that no cancellations occur amongst the LFV parameters ($|g_{\rm lfv}| \simeq |\tilde{h}^{}_{\mu e}\tilde{h}^*_{ee}|$), one expects ${\rm BR}_{\mueee}$ to be roughly two orders of
magnitude larger than ${\rm R}^A_{\mue}$ for ${\cal O}({\rm TeV})$ Higgs triplet masses \cite{Cirigliano:2004mv}. Thus in this simplified case the limits on $\mueee$ will confine the model parameter
space the most.

However, with right-handed neutrinos around the TeV scale the left-right mixing could be enhanced, so that the usual type I seesaw contribution to LFV processes should also be considered. Those have
been calculated in Refs.~\cite{Bilenky:1977du,Cheng:1980tp,Ilakovac:1994kj,Ioannisian:1999cw,Lavoura:2003xp,Pilaftsis:2005rv,Dinh:2012bp,Alonso:2012ji,Tello_thesis:2012}. Since the LRSM is effectively
a type~I+II seesaw model one needs to take into account LFV processes mediated by both heavy neutrinos and Higgs triplets, effectively allowing for interference between different amplitudes.
Ref.~\cite{Tello_thesis:2012} has presented the full expressions for $\muegam$; we include type I seesaw terms in the expressions for $\mueee$ and $\mue$ conversion, in the former case including
possible interference between loop and tree level diagrams. Detailed expressions for the decay widths including form factors and loop functions can be found in the appendix; we summarize the most
constraining processes here. In our 
parameter scans in the type I dominance case we take into account all relevant contributions.

It turns out that the most important constraints on the mixing $S \simeq M_D/M_R$ come from $\muegam$ and $\mue$ conversion. In both cases the constraint is roughly 
\begin{equation}
 S^*_{\mu i}S^{}_{ei} {\cal F}(x_i) \simeq S^*_{\mu i}S^{}_{ei} \ls 10^{-5}\, ,
\label{eq:most_stringent_bound}
\end{equation}
where we take the loop function ${\cal F}(x_i)$ to be of order one. This approximation is not always valid for very large right-handed neutrino masses, in which case ${\cal F}(x_i) \simeq
\ln(M_i^2/\mwl^2)$, but since the mixing scales with $1/M_i$ the rate will vanish in the decoupling limit \cite{Alonso:2012ji}. The loop-suppressed decay rate of $\mueee$ (with heavy neutrinos
exchanged) depends on the same parameters as $\mue$ conversion, but the limits are weaker in this case: the bound ${\rm BR}_{\mueee} < 1.0 \times 10^{-12}$ can be roughly translated into $S^*_{\mu i}
S^{}_{e i} \ls 10^{-3}$. These constraints come from diagrams with left-handed currents and left-right mixing, i.e. the terms proportional to $S^2$ in Eqs.~\eqref{eq:dipole_form_facts},
\eqref{eq:fgamm_form}, \eqref{eq:z_form_fact} and \eqref{eq:box_form_facts_mueee}, so that there is no other dependence on the heavy particle masses besides from the loop functions. Another
interesting constraint comes from $\muegam$ diagrams in which 
gauge bosons mix: the chirality flip occurs within the loop, leading to a direct dependence on the Dirac mass matrix instead of the muon mass [Eq.~\eqref{eq:dipole_form_facts}], in a similar way to
the mixed diagrams in $\obb$ (see also Refs.~\cite{He:2002pva,Lavoura:2003xp,Tello_thesis:2012}). This enhances the contribution of mixed diagrams to $\muegam$ by a factor $SM_R/m_\mu\simeq
M_D/m_\mu$, so that the product of the mixing angle $\xi$ and the $\mu e$ element of the Dirac mass matrix is constrained to be
\begin{equation}
 |M^*_D|_{\mu e} \left(\frac{\xi}{10^{-5}}\right) \ls 0.2 \ {\rm GeV}\, .
\end{equation}
In addition, the experimental limit of $|d_e| < 10^{-27}\, e$~cm \cite{Hudson:2011zz} on the electric dipole moment of the electron [see Eq.~\eqref{eq:edipole}] constrains the $e e$ element to be
roughly 
\begin{equation}
 {\rm Im}\left\{[M_D]_{e e}e^{i\alpha}\right\} \left(\frac{\xi}{10^{-5}}\right) \ls 0.02 \ {\rm GeV}\, ,
\end{equation}
which also depends on the phase $\alpha$. These limits effectively constrain the $\eta$-diagram in Fig.~\ref{fig:obb_LR_eta}.

One might also expect large left-right mixing to allow loop-suppressed (type~I) contributions to $\mueee$ to compete with the tree level triplet (type~II) contribution. The full expression is given in
Eq.~\eqref{eq:brmueee_LL_full}, and the condition for comparable magnitudes of type~I and type~II contributions is roughly
\begin{equation}
S^*_{\mu i}S^{}_{ei} \simeq 0.1 \left(\frac{5\ {\rm TeV}}{m_{\delta^{++}}}\right)^2 \left(\frac{\left|M^{}_{\mu e}M^*_{ee}\right|}{\mwr^2}\right),
\end{equation}
assuming $\mdl = \mdr \equiv m_{\delta^{++}}$. Thus for TeV-scale $W_R$ the bound on $S^2$ in Eq.~\eqref{eq:most_stringent_bound} means that one needs right-handed neutrinos around the electroweak
scale for the type~I loop contribution to be competitive in $\mueee$ decay. 

\subsection{Collider physics} \label{subsect:collider}

Before concentrating on the $\obb$ amplitudes we briefly discuss the role of collider physics in studying the LRSM. Collider searches provide a complementary probe of the parameter space of the LRSM:
the right-handed $W$ boson and right-handed neutrinos can be produced in $pp$ collisions at the LHC via~\cite{Keung:1983uu}
\begin{equation}
 pp \to W_R + X \to N_\ell + \ell + X\, , \quad (\ell = e,\mu),
\end{equation}
followed by the decay into like-sign dileptons and two jets, i.e. 
\begin{equation}
 W_R \to \ell_1N_\ell\to\ell_1\ell_2W_R^*\to\ell_1\ell_2qq'\to\ell_1\ell_2jj\, ,
\end{equation}
which for the $\ell=e$ case is equivalent to the $\obb$ diagram in Fig.~\ref{fig:obb_LR_NR}. The CMS collaboration looked for this signature in both 7~TeV~\cite{CMS:2012zv} and
8~TeV~\cite{CMS:2012uwa} data, where the integrated luminosity was 5.0 fb$^{-1}$ and 3.6 fb$^{-1}$, respectively. Their analysis was simplified by assuming negligible mixing ($\xi\simeq 0$) between
gauge bosons and between heavy neutrino mass eigenstates ($V \simeq \mathbb{1}$), so that the final states are either both electrons or both muons. ATLAS studied the same process with 2.1~fb$^{-1}$ of
data from 7~TeV collisions \cite{ATLAS:2012ak}, and in addition examined the case of maximal mixing between the first two heavy neutrino mass eigenstates.

\begin{figure}[t]
 \centering
 \includegraphics[width=0.5\textwidth]{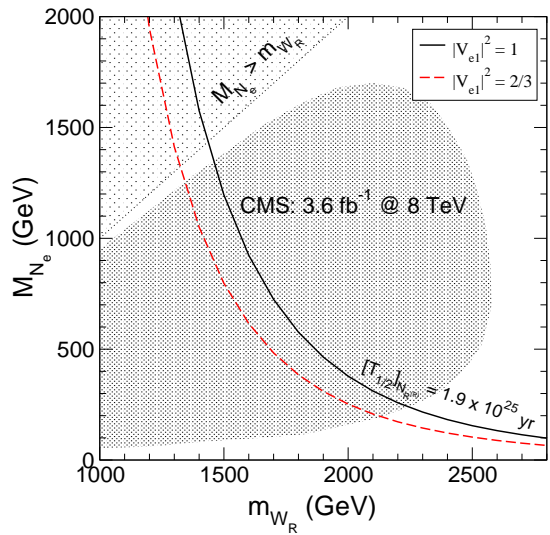}
 \caption{Comparison of the limits in $M_{N_e}-\mwr$ parameter space from CMS and from the KamLAND-Zen limit on $\obb$. The limit of $1.9 \times 10^{25}$~yrs on the $\obb$ half-life of $^{136}$Xe
means that all points to the left of the solid black line (dashed red line) are excluded, for $|V_{ei}|^2 = 1$ ($|V_{ei}|^2 = 2/3$), where we assume only heavy neutrinos contribute to $\obb$,
i.e.~only $[T_{1/2}^{0\nu}]_{N^{(R)}_R}$. The shaded region is excluded by CMS at 95\% C.L. \cite{CMS:2012uwa}.}
\label{fig:mne_mwr_CMS}
\end{figure}

As a simple illustration of the complementarity of the different data sets we plot the limits from the latest CMS data as well as from the KamLAND-Zen $\obb$ experiment~\cite{Gando:2012zm} in the
$M_{N_e}-\mwr$ parameter space in Fig.~\ref{fig:mne_mwr_CMS}, using two different values for the mixing $V_{e1}$. Here one assumes that only one heavy neutrino flavour $N_e \simeq N_1$ is accessible,
so that the LNV parameter in Eq.~\eqref{eq:eta_RNR_def} simply becomes $|\eta^R_{N_R}| = m_p(\mwl/\mwr)^4|V^*_{e1}|^2/M_1$. 

It is also possible to probe the couplings $h_{\alpha\beta}$ of Higgs triplets to leptons [see Eq.~\eqref{eq:lag_full_lep}] with collider searches. The latest results from ATLAS \cite{ATLAS:2012hi}
give the exclusion limits $m_{\delta^{\pm\pm}_L} > 409$~GeV and $m_{\delta^{\pm\pm}_R} > 322$~GeV for $e^\pm e^\pm$ final states and assuming a branching ratio of 100\% to each final state. In order
to compare these results to the $\obb$ bounds one needs to take into account the other decay modes of doubly-charged Higgs scalars into gauge bosons and singly-charged scalars. An analysis in this
direction was performed in Ref.~\cite{Melfo:2011nx}, and the results depend largely on the mass spectrum of the different components of the Higgs triplets $\Delta_{L,R}$.

\section{\texorpdfstring{$\obb$ amplitudes in the seesaw limits}{0nubb amplitudes in the seesaw limits}} \label{sec:amp_calcs}

In the most general case the light neutrino mass matrix
\begin{equation}
 m_\nu = M_L - M_DM_R^{-1}M_D^T\, ,
\end{equation}
receives contributions from  [see Eq.~\eqref{eq:mnu_def}] both the left-handed triplet (type II seesaw) and the heavy right-handed neutrinos (type I seesaw), making quantitative studies of the $\obb$
amplitudes difficult. Here we focus on the two extreme cases of type II and type I dominance; a complete study is beyond the scope of this work. In the former case one sets the Dirac Yukawa couplings
to zero, in the latter one assumes that the triplet VEV vanishes, i.e., $v_L = 0$. The simpler case of type II seesaw dominance is dealt with first; this was first studied in
Ref.~\cite{Tello:2010am} and further examined in Ref.~\cite{Chakrabortty:2012mh}.

\subsection{Type II seesaw dominance} \label{subsec:typeII}

With the approximations mentioned above, the lifetime in the limit of type II dominance is
\begin{equation}
 \left[T_{1/2}^{0\nu}\right]_{\rm type~II}^{-1} = G^{0\nu}_{01}\left\{\left|{\cal M}^{0\nu}_\nu\right|^2|\eta_\nu|^2+\left|{\cal M}_N^{0\nu}\right|^2|\eta^R_{N_R}+\eta_{\delta_R}|^2\right\};
\label{eq:half-life_typeII}
\end{equation}
by neglecting all Dirac Yukawa couplings we drop all terms proportional to $M_D$, i.e., those with left-right mixing. We are left with only heavy neutrino [Fig.~\ref{fig:obb_LR_NR}] and triplet
exchange [Fig.~\ref{fig:tripletRR}] in addition to the standard diagram [Fig.~\ref{fig:obb_LR_standard}] (the amplitude ${\cal A}^L_{N_R}$ also vanishes, being proportional to $M_D$). As discussed
above, the interference term is suppressed, since the final state electrons in Fig.~\ref{fig:obb_LR_standard} are left-handed whereas those in Fig.~\ref{fig:obb_LR_NR} are right-handed. 

\begin{figure}[tpb]
 \centering
 \includegraphics[angle=270,width=0.9\textwidth]{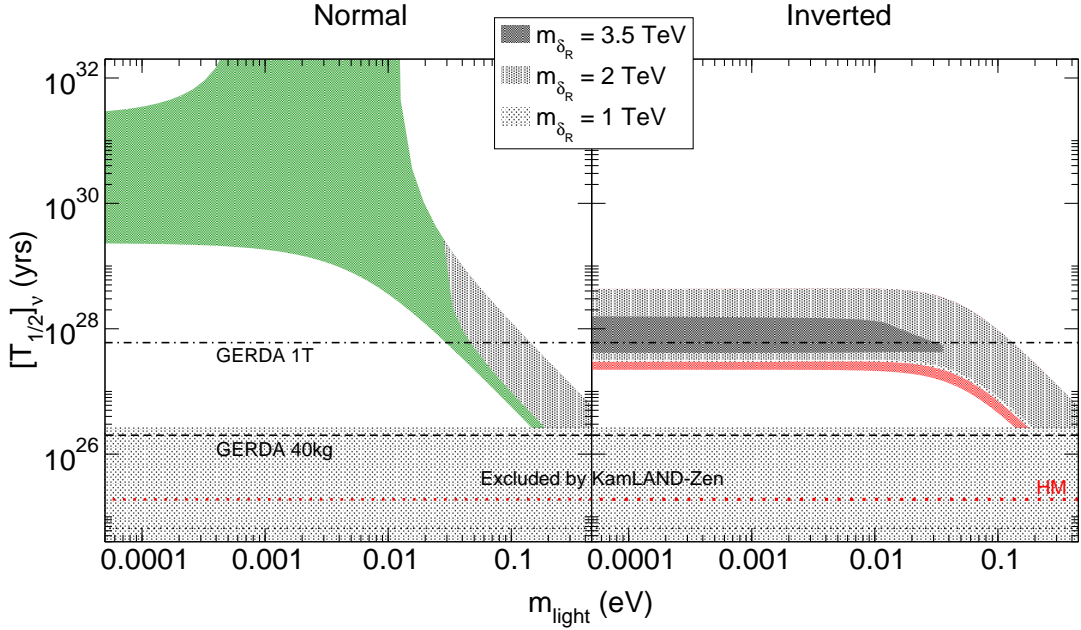}
 \caption{Same as Fig.~\ref{fig:standard_lifetime_Ge}, with the grey shaded areas forbidden by the LFV constraint ${\rm BR}_{\mu \to 3e} \ls 10^{-12}$, for different values of the Higgs triplet mass,
with $\mwr = 3.5$~TeV and the heaviest right-handed neutrino $M_{\rm heavy}=500$~GeV.}
\label{fig:standard_lifetime_Ge_LFV}
\end{figure}

In the case of type II dominance, the right-handed neutrino mass matrix can be expanded as \cite{Akhmedov:2006de}
\begin{equation}
 M_R^{\rm type\ II} \simeq \frac{v_R}{v_Le^{i\theta_L}}m_\nu + \kappa_+^2h_D m_\nu^{-1} h_D^T - \kappa_+^4\frac{v_Le^{i\theta_L}}{v_R}(h_Dm_\nu^{-1}h_D) m_\nu^{-1}(h_D m_\nu^{-1}h_D)^T + \ldots,
\end{equation}
and since we neglect Yukawa couplings ($h_D \approx 0$),
\begin{equation}
 M_R = \frac{v_R}{v_Le^{i\theta_L}}m_\nu\, ,
\label{eq:typeII_dom_relation}
\end{equation}
which simplifies the analysis considerably: the light and heavy neutrino spectra are proportional to each other, and $V = U$, up to an overall complex phase. In addition, both $U$ and $V$ become
unitary in the limit that $M_D = 0$ [cf.~Eq.~\eqref{eq:mix_matrices}]. These assumptions were used in Ref.~\cite{Tello:2010am} to quantify the heavy neutrino contribution to \onbb; the triplet
contribution to $\obb$ was neglected since the constraint from $\mueee$ leads to  $M_R/m_{\delta_R} \ll 1$ over a large part of parameter space. It is however useful to consider the different
contributions in more detail, since there are areas of parameter space where the triplet contribution gives interesting effects (see also Ref.~\cite{Chakrabortty:2012mh}). Here we calculate the
relevant lifetimes in each case and show explicit regions in parameter space where the limit from ${\rm BR}_{\mueee}$ comes into play. Replacing $V$ with $U$ in Eq.~\eqref{eq:eta_RNR_def}, the
dimensionless LNV parameter corresponding to
heavy neutrino exchange with right-handed currents ($\propto [M_R^{-1}]_{ee}$) can now be written as
\begin{align}
[\eta^R_{N_R}]_{\rm NO} &= m_p\left(\frac{\mwl}{\mwr}\right)^4\left(\frac{m_3}{m_1}|U_{e1}|^2+\frac{m_3}{m_2}|U_{e2}|^2e^{-i\alpha}+|U_{e3}|^2e^{-i\beta}\right)\frac{1}{M_3}\, , \\
[\eta^R_{N_R}]_{\rm IO} &= m_p\left(\frac{\mwl}{\mwr}\right)^4\left(\frac{m_2}{m_1}|U_{e1}|^2+|U_{e2}|^2e^{-i\alpha}+\frac{m_2}{m_3}|U_{e3}|^2e^{-i\beta}\right)\frac{1}{M_2}\, ,
\end{align}
for normal and inverted ordering, respectively, where $\alpha$ and $\beta$ are Majorana phases. Similarly, the branching ratio for $\mueee$ in Eq.~\eqref{eq:brmueee_tree} depends on the product of the
$ee$ and $\mu e$ elements of $\tilde{h} = M_R/\mwr$, with
\begin{align}
 \left[M_R\right]^{\rm NO}_{\sigma \rho} &= \left(\frac{m_1}{m_3}U_{\sigma 1}U_{\rho 1}  + \frac{m_2}{m_3}U_{\sigma 2}U_{\rho 2}e^{i\alpha}  + U_{\sigma 3}U_{\rho 3}e^{i\beta} \right)M_3\, , \\
 \left[M_R\right]^{\rm IO}_{\sigma \rho} &= \left(\frac{m_1}{m_2}U_{\sigma 1}U_{\rho 1}  + U_{\sigma 2}U_{\rho 2}e^{i\alpha}  + \frac{m_3}{m_2}U_{\sigma 3}U_{\rho 3}e^{i\beta} \right)M_2\, .
\end{align}
We assume $\mdl=\mdr$ in what follows.

\begin{figure}[tpb]
 \centering
  \subfigure[]{\label{fig:thalf_R}
  \includegraphics[angle=270,width=0.7\textwidth]{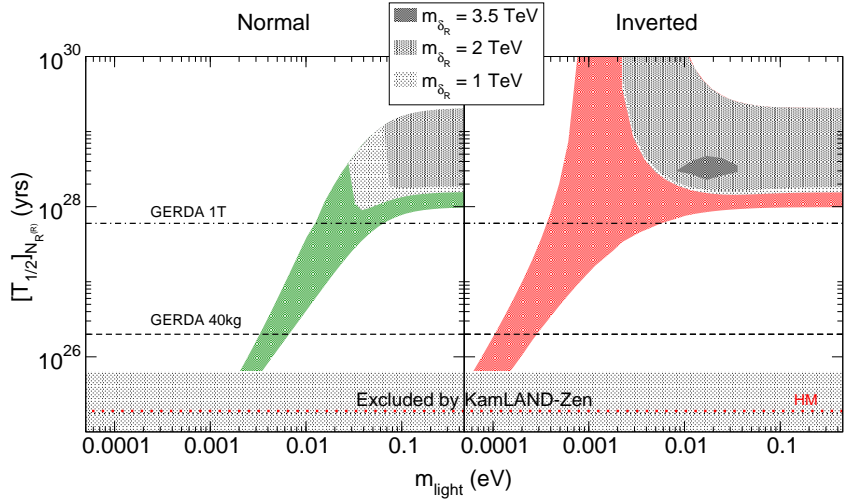}}
  \subfigure[]{\label{fig:thalf_delR}
  \includegraphics[angle=270,width=0.7\textwidth]{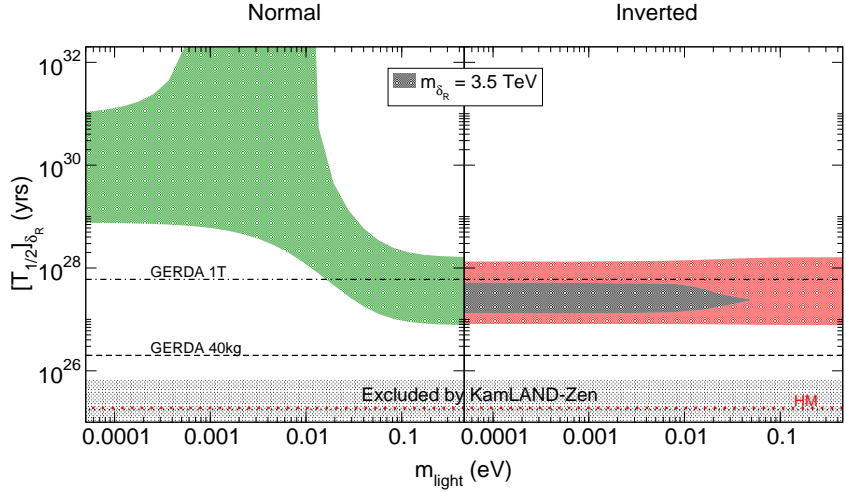}}
 \caption{The contribution to the $\obb$ half-life of $^{76}$Ge from (a) heavy right-handed neutrinos and (b), right-handed Higgs triplets, plotted against the lightest light neutrino mass, with $\mwr
= 3.5$~TeV and $M_{\rm heavy}=500$~GeV. In plot (a) the grey shaded regions are excluded by LFV constraints, for different values of $\mdr$, in plot (b) $\mdr=\mwr=3.5$~TeV. Experimental limits are
explained in the caption of Fig.~\ref{fig:standard_lifetime_Ge}.}
 \label{fig:thalf_R_delR}
\end{figure}

Following Ref.~\cite{Tello:2010am}, by fixing $\mwr = 3.5$~TeV and the heaviest right-handed neutrino mass $M_{\rm heaviest} = 500$~GeV, the three contributions can be plotted against the lightest
light neutrino mass (see Figs.~\ref{fig:standard_lifetime_Ge_LFV} and \ref{fig:thalf_R_delR}). It is clear that the right-handed contribution $[T_{1/2}]^{-1}_{N_R^{(R)}}$ [Fig.~\ref{fig:thalf_R}] is
proportional to the inverse of $M_R$, whereas the triplet contribution $[T_{1/2}]^{-1}_{\delta_R}$ [Fig.~\ref{fig:thalf_delR}] is proportional to $M_R$, and looks similar to the standard lifetime
$[T_{1/2}]^{-1}_\nu$ (Fig.~\ref{fig:standard_lifetime_Ge_LFV}), since $m_\nu \propto M_R$ in the type II limit. For $[T_{1/2}]^{-1}_{N_R^{(R)}}$, the inverted ordering can have infinite lifetime (zero
effective mass), whereas the normal ordering cannot, so that the roles are reversed with respective to the standard case. In each plot we indicate the regions excluded by the limit on $\mueee$ for
different values of $\mdr$: in 
the normal hierarchy the constraint only comes into play when the lightest mass is larger than about 0.01~eV, whereas in the inverted hierarchy the whole parameter space is affected\footnote{Our
results agree with Fig.~2 of Ref.~\cite{Tello:2010am}, which shows that $M_{\rm heavy}/\mdr \ls 0.1$ in the inverted ordering for all light neutrino masses, which in our case would correspond to $\mdr
= 5$~TeV.}. In the case of the light neutrino and triplet contributions, the only areas still allowed correspond to the largest possible value of $\mee$, i.e., when both Majorana phases are close to
zero.

\begin{figure}[t]
 \centering
  \includegraphics[angle=270,width=0.9\textwidth]{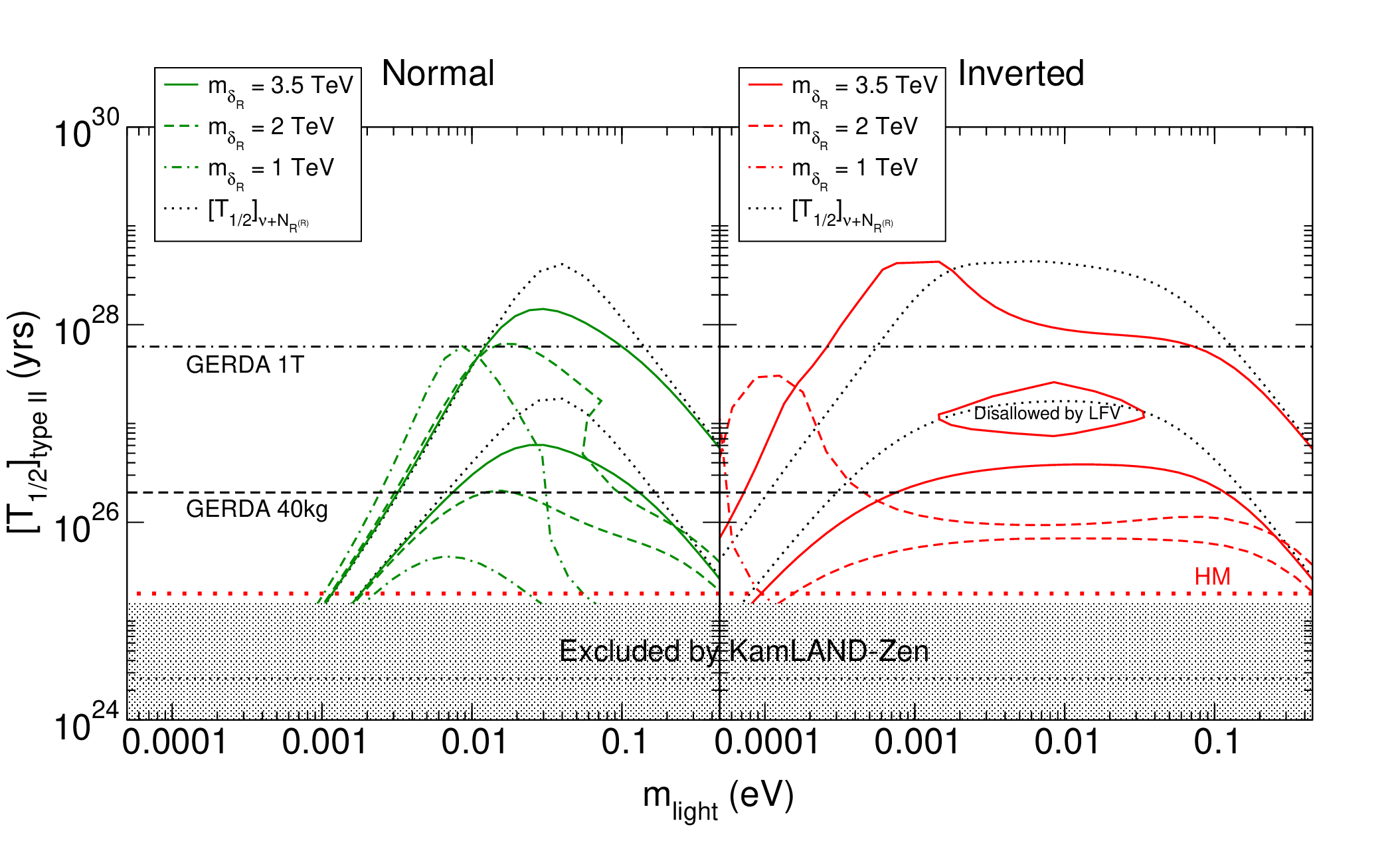}
 \caption{The total $\obb$ half-life of $^{76}$Ge including light neutrino, heavy neutrino and triplet contributions, plotted against the lightest light neutrino mass, with $\mwr = 3.5$~TeV and
$M_{\rm heavy}=500$~GeV. The solid, dashed and dashed-dotted lines show the allowed regions that satisfy ${\rm BR}_{\mueee}\leq 10^{-12}$ for $\mdr$ equal to $1$, $2$ and $3.5$~TeV respectively; the
black dotted lines enclose the regions allowed if one neglects the triplet contribution and LFV constraints. Experimental limits are explained in the caption of Fig.~\ref{fig:standard_lifetime_Ge}.}
 \label{fig:thalf_tot_II}
\end{figure}

Figure~\ref{fig:thalf_tot_II} shows the total half-life, with all three contributions included. The chosen value of $\mdr$ affects not only the LFV constraint but also the resulting half-life, due to
the dependence of the triplet contribution on this quantity [Eq.~\eqref{eq:amp_triplet_R}]. The black dotted lines show the half-life without the triplet contribution, and it is evident that the
addition of the triplet part can shorten the half-life by several orders of magnitude, bringing it within reach of the GERDA experiment. There also exist regions where the lifetime can be longer, due
to cancellations between the $\eta^R_{N_R}$ and $\eta_{\delta_R}$ contributions. The key point  here is that the triplet contribution can still be allowed for certain values of the Majorana phases,
even with the LFV constraint, thus enhancing the total amplitude for $\obb$. This enhancement obviously depends on the triplet mass, so that if $\mdr \gs 5$~TeV we recover the results of
Ref.~\cite{Tello:2010am}.

\subsection{Type I seesaw dominance} \label{subsec:typeI}

In the limit of type I seesaw dominance all the terms in Eq.~\eqref{eq:half-life_simp} must be considered (we neglect the small contribution from $\eta_{\delta_L}$, as discussed above). This leaves us
with six contributing diagrams: $(i)$~``standard'' light neutrino exchange ($\eta_\nu$); $(ii)$~heavy neutrino exchange with left-handed currents ($\eta^L_{N_R}$); $(iii)$~heavy neutrino exchange with
right-handed currents ($\eta^R_{N_R}$); $(iv)$~light neutrino exchange via the $\lambda$-diagram ($\eta_\lambda$); $(v)$~light neutrino exchange via the $\eta$-diagram ($\eta_\eta$) and
$(vi)$~right-handed triplet exchange ($\eta_{\delta_R}$). There are also interference terms [see Eq.~\eqref{eq:half-life_full}], and distinguishing the different contribution becomes difficult.
Although most studies focus on the standard diagram and those with heavy neutrinos, the contributions $(iv)$ and $(v)$ can actually be significant, as we have shown in the rough estimates above. These
have been studied in for example Refs.~\cite{Hirsch:1996qw,Pas:1999fc}.

\subsubsection{Parameterizing the relative magnitudes} \label{sect:casas-ibarra}

In order to quantify the six contributions one needs more information about the right-handed sector, specifically the right-handed mixing matrix $V_R$ and the mass spectrum $M_i$ ($i=1,2,3$) of
right-handed neutrinos. The right-handed mass matrix $M_R$ appears in the amplitudes ${\cal A}^L_{N_R}$, ${\cal A}^R_{N_R}$, ${\cal A}_{\delta_R}$, ${\cal A}_\lambda$ and ${\cal A}_\eta$, and in the
case of type I seesaw dominance can be expanded as
\begin{equation}
 M_R^{\rm type\ I} = \kappa_+^2 h_D^T m_\nu^{-1} h_D + \kappa_+^4\frac{v_Le^{i\theta_L}}{v_R}(h_Dm_\nu^{-1}h_D)^Tm_\nu^{-1}(h_Dm_\nu^{-1}h_D) + \ldots
\end{equation}
The leading term is a matrix product containing the unknown Dirac mass matrix, so that the simple relations in Eq.~\eqref{eq:typeII_dom_relation} no longer hold and one needs a different approach. The
authors of Ref.~\cite{Chakrabortty:2012mh} simplify the analysis by assuming that $(i)$ the Dirac mass matrix is diagonalized by $V_R$ and $(ii)$ the three Dirac Yukawas are equal.  This scenario is
very restrictive; another approach would be to insert an ansatz for the matrix of Dirac Yukawa couplings $h_D$. Often one uses the condition $M_u \simeq M_D = \kappa_+h_D$, which holds at the GUT
scale in $SO(10)$ models~\cite{Hosteins:2006ja}. 

More generally, the Dirac mass matrix can be parameterized using the so called top-down or ``$V_L$--parameterization''
\begin{equation}
 M_D = U_L^\dagger \tilde{M}_D U_R\,,
\label{eq:md_vl_param}
\end{equation}
where $U_L$ and $U_R$ are arbitrary unitary matrices and $\tilde{M}_D = \kappa_+\,{\rm diag}(h_1,h_2,h_3)$. In the LRSM type I case, $M_D$ has 18 parameters and $M_R$ has 12 parameters, so that the
left-right mixing $M_D M_R^{-1}$ depends on 30 parameters, making it difficult to learn anything from a parameter scan. If we assume a discrete parity (charged conjugation) symmetry, then $M_D$
becomes hermitian (symmetric) thus reducing the number of parameters by 6. However, it is still numerically difficult to find Dirac mass matrix structures that give large enough left-right mixing. One
way is to start from a specific matrix structure in $M_D$ that gives zero neutrino masses, and introduce small perturbations (see Refs.~\cite{DeGouvea:2007uz,Kersten:2007vk}).

An alternative method is to go to the basis where $M_D$ is ``diagonal'', so that the light neutrino mass matrix is given by
\begin{equation}
 m'_\nu = -\tilde{M}_D{M'_R}^{-1}\tilde{M}_D\, ,
\label{eq:mnu_mddiag}
\end{equation}
with ${M'_R}^{-1} = U_RM_R^{-1}U_R^T$. In essence one has rotated the left-handed neutrino fields by $U_L$ [cf.~Eq.~\eqref{eq:md_vl_param}]. After diagonalizing $m'_\nu$ by the unitary matrix $X_L$,
i.e.~$m'_\nu=X_L\tilde{m}_\nu X_L^T$, the neutrino mass matrix in the flavour basis is
\begin{equation}
 m_\nu = -V_\nu X_L^\dagger\left(\tilde{M}_D{M'_R}^{-1}\tilde{M}_D\right) X_L^*V_\nu^T \equiv -U_L^\dagger\left(\tilde{M}_D{M'_R}^{-1}\tilde{M}_D\right)U_L^*\, , \label{eq:mnuflav}
\end{equation}
where $V_\nu$ is the light neutrino mixing matrix [Eq.~\eqref{eq:mnu_MR_def}] defined by $V_\nu \equiv U_L^\dagger X_L$. Numerically, this means one needs only fit the mass eigenvalues after
diagonalizing Eq.~\eqref{eq:mnu_mddiag}, decoupling the PMNS mixing parameters\footnote{This approach is discussed in Ref.~\cite{Casas:2006hf}.}. The authors of Ref.~\cite{Mitra:2011qr} used this
approach to find matrix structures that could enhance the amplitude for double beta decay mediated by heavy sterile neutrinos (${\cal A}^L_{N_R}$), albeit without right-handed currents. In our case
those same structures will also enhance the amplitudes for the $\lambda$- and $\eta$-diagrams and influence the LFV branching ratios. However, one cannot recover the non-trivial mixing $V_R$ in the
right-handed sector simply by diagonalizing ${M'_R}^{-1}$. Defining ${M'_R}^{-1} = X_R^*\tilde{M}_R^{-1}X_R^\dagger$ means that
\begin{equation}
 V_R = U_R^TX^{}_R\, ,
\end{equation}
so that the only way to find $V_R$ is to invoke the symmetry (hermiticity) of $M_D$, which gives $U_R = U_L^*$ ($U_R = U_L$). The right-handed mixing is then
\begin{equation}
 V_R = U_L^\dagger X_R = V_\nu X_L^\dagger X_R \quad {\rm or} \quad V_R = U_L^TX_R = V_\nu^*X_L^TX_R\, ,
\end{equation}
whereas the left-right mixing (in the flavour basis) is
\begin{equation}
 M^{}_DM_R^{-1} = U_L^\dagger\tilde{M}_D{M'_R}^{-1}U^{}_L \quad {\rm or} \quad M^{}_DM_R^{-1} = U_L^\dagger\tilde{M}_D{M'_R}^{-1}U^*_L\, ,
\label{eq:lrmixing_md_diag}
\end{equation}
for symmetric or hermitian $M_D$, respectively. The expression [cf.~Eq.\eqref{eq:md2mr3}] characterizing the diagram with heavy neutrinos and left-handed currents is
\begin{equation}
M^{}_DM_R^{-1}{M_R^{-1}}^*M_R^{-1}M_D^T = U_L^\dagger \tilde{M}^{}_D{M'_R}^{-1}{{M'_R}^{-1}}^*{M'_R}^{-1}\tilde{M}^{}_DU_L^*\, .
\end{equation}
The corrected forms of $U$ and $V$ used for calculating $\obb$ amplitudes and LFV branching ratios can be found from Eq.~\eqref{eq:mix_matrices}, but in our case the terms second order in $R \simeq
M^{}_DM_R^{-1}$ make little difference.

The main point is that there are certain regions of parameter space which allow for large left-right mixing while still keeping the light neutrino masses small enough, since the matrix
structures allow for cancellations. One could regard this as a fine-tuned scenario; on the other hand it is obvious that there is enough freedom in parameter space to allow for it. For completeness
we note that it is possible to scan the entire allowed parameter space using the orthogonal parameterization \cite{Casas:2001sr}, where the Dirac mass matrix is written as\footnote{Note that in the
left-right model we cannot rotate to a basis where $M_R$ is diagonal without affecting the right-handed charged current.} $M_D = i\, V_\nu {\tilde{m}_\nu}^{1/2}O{\tilde{M}_R}^{1/2}V_R^T$, with $OO^T =
O^TO = \mathbb{1}$ and the diagonal matrices $\tilde{m}_\nu = {\rm diag}(m_1,m_2,m_3)$ and $\tilde{M}_R = {\rm diag}(M_1,M_2,M_3)$. 

It has also been shown \cite{Nemevsek:2012iq} that if the Dirac mass matrix
is symmetric, there are only $2^3=8$ discrete solutions to the seesaw equation, given by $M_D = i\sqrt{m_\nu M_R^{-1}}M_R$, so that the $O$ matrix in the orthogonal parameterization is given by $O =
{\tilde{m}_\nu}^{-1/2}V_\nu^\dagger\left(m^{}_\nu M_R^{-1}\right)^{1/2}V_R{\tilde{M}_R}^{1/2}$. However, one still has a large number of unknown parameters in the right-handed sector, 
and the $O$-matrix approach does not allow one to define a symmetric or hermitian Dirac mass matrix in a simple way. We have checked that it is possible to use the method of
Ref.~\cite{Mitra:2011qr} (described above) to obtain large left-right mixing solutions that are consistent with this formalism. In that case half of the eight solutions give large mixing, whereas the
other half give small mixing.

\subsubsection{Numerical example}

In the most general case, one should solve the condition $M_DM_R^{-1}M_D^T = 0$ in order to find solutions with large mixing, and it turns out that in the basis in Eq.~\eqref{eq:mnu_mddiag}
this equates to \cite{Mitra:2011qr}
\begin{equation}
 \tilde{M}_D \propto {\rm diag}(0,0,1) \quad {\rm and} \quad M'_R \propto \begin{pmatrix} 0 & 0 & 1 \\ 0 & 1 & 1 \\ 1 & 1 & 1 \end{pmatrix}.
\end{equation}
Inserting small parameters instead of zeros leads to non-zero light neutrino masses, with the spectrum depending on any hierarchies introduced in $\tilde{M}_D$ and $M_R$. One particular example (from
Ref.~\cite{Mitra:2011qr}) is
\begin{equation}
 \tilde{M}_D = \kappa^+{\rm diag}(a \epsilon^2,b \epsilon,c), \quad {M'_R}^{-1} \simeq M^{-1}\begin{pmatrix} d & e & f \\ \cdot & g & h \epsilon \\ \cdot & \cdot & j \epsilon^2 \end{pmatrix},
\label{eq:texture_example}
\end{equation}
which leads to nonzero lightest neutrino mass. With all coefficients $a$, $b$, $c$ etc.~of order one one needs $|\epsilon| = {\cal O}(10^{-6})$ in order to get the correct mass for active neutrinos
with the matrix textures in Eq.~\eqref{eq:texture_example}. Inverting ${M'_R}^{-1}$ would give a matrix with small $(1,1)$, $(1,2)$ and $(2,1)$ entries, but since ${M'_R}^{-1} =
\left(U_RM_R^{-1}U_R^T\right)$, the matrix $M_R$ can have large entries everywhere, which can enhance the LFV amplitudes. This is simply a manifestation of the fact that one cannot go to a basis where
the right-handed neutrinos are diagonal without affecting the right-handed current, which is different to the conventional case. For our parameter scans we set $\mwr=3.5$~TeV and $\mdr = 5$~TeV and
vary the gauge boson mixing angle in the range $10^{-8} \leq \xi \leq 10^{-6}$ , otherwise it would be difficult to evade the constraints from $\muegam$. The magnitudes of the complex parameters $a$,
$b$, $c$ etc.~are varied in the range $[0.1,1.0]
$, and $|\epsilon|$ in the range $[10^{-12},10^{-5}]$. The phases are taken to be between $0$ and $2\pi$, and $\kappa^+ = 174$~GeV and $M = 1$~TeV are fixed. From Eqs.~\eqref{eq:md_ml_mr} and
\eqref{eq:mw1_2} the relation $M_R = \frac{2}{g}\mwr h \simeq 3 \mwr h$ holds, which we used to check perturbativity of the coupling $h$. An explicit numerical example is given in
Appendix~\ref{sect:explicit_example}.

\begin{figure}[t]
 \centering
 \includegraphics[angle=270,width=0.9\textwidth]{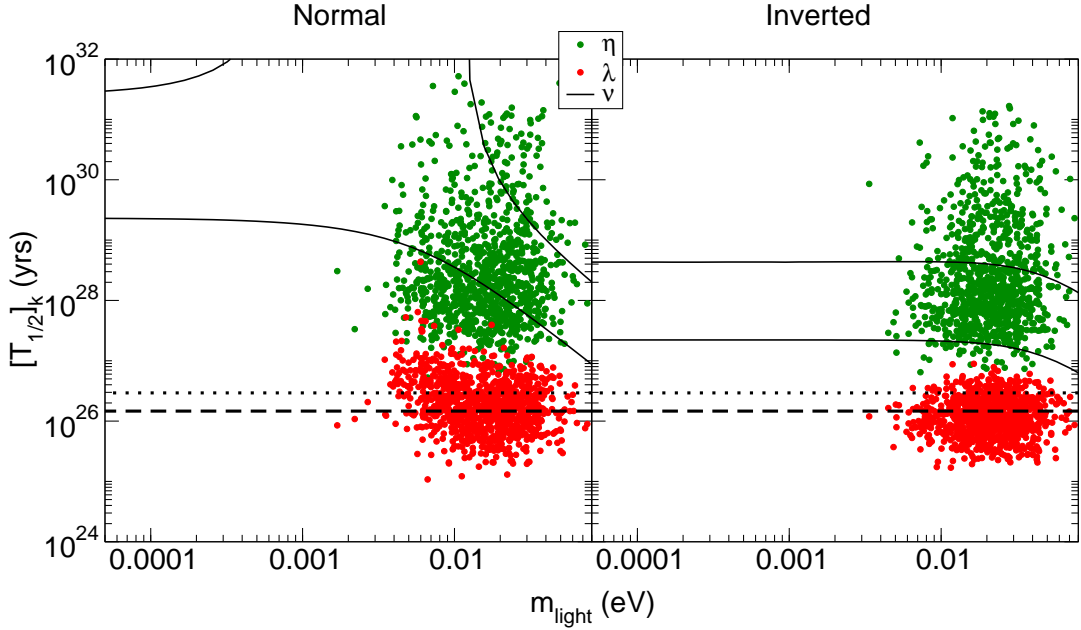}
 \caption{Contribution to the $\obb$ half-life of $^{76}$Ge from the $\lambda$- and $\eta$-diagrams plotted against the lightest light neutrino mass, for symmetric $M_D$. The standard contribution is
indicated by the region outlined in black, and the dashed and dotted horizontal line correspond to the limits from Eqs.~\eqref{eq:etal} and Eq.~\eqref{eq:etaeta}.}
\label{fig:lam_eta_nu_mlight}
\end{figure}

\begin{figure}[t]
 \centering
 \includegraphics[angle=270,width=0.9\textwidth]{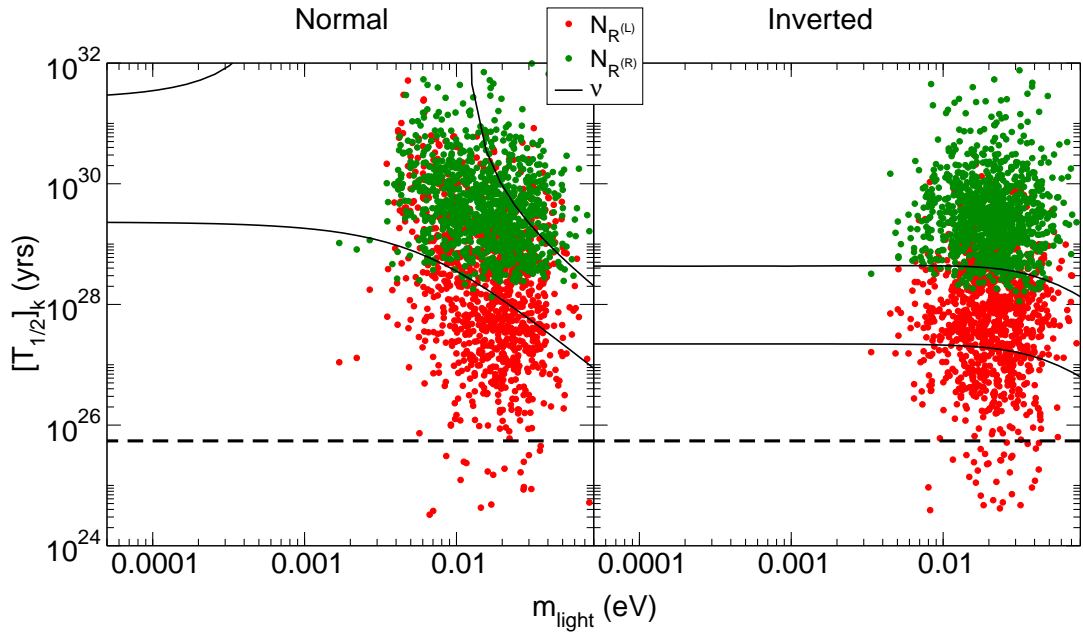}
 \caption{Contribution to the $\obb$ half-life of $^{76}$Ge from heavy right-handed neutrinos, with left- and right-handed currents (${\cal A}^{L,R}_{N_R}$), for symmetric $M_D$.  The standard
contribution is indicated by the region outlined in black, and the dashed horizontal line corresponds to the limit from Eq.~\eqref{eq:etalnr_lim}.}
\label{fig:LRNR_nu_mlight}
\end{figure}

One expects the different half-life contributions to have similar orders of magnitude, since we are exploring the fine-tuned region, so that the amplitudes ${\cal A}^L_{N_R}$, ${\cal A}_\lambda$ and
${\cal A}_\eta$, which all depend on the left-right mixing, are enhanced. We plot the halflives for the amplitudes ${\cal A}_\lambda$ and ${\cal A}_\eta$ in Fig.~\ref{fig:lam_eta_nu_mlight} and the
halflives corresponding to heavy neutrino exchange, i.e.~the amplitudes ${\cal A}^L_{N_R}$, ${\cal A}^R_{N_R}$ and ${\cal A}_{\delta_R}$ in Fig.~\ref{fig:LRNR_nu_mlight}, in both cases for a symmetric
Dirac mass matrix. In each case the usual light neutrino contribution is shown for comparison, and one can see that there are regions of parameter space in which the $\lambda$ and $\eta$ contributions
dominate over the light neutrino contribution. Remarkably the $\eta$ contribution can still be sizeable, even with such small values of $\xi$: this is largely due to the larger value of the matrix
element ${\cal M}^{0\nu}_\eta$ (cf.~Table~\ref{table:matrix_elements}). The lightest mass could be smaller if the parameters $a$, $b$, $c$ were allowed to be smaller than 0.1, although in the normal
ordering case the LFV constraints in general favour larger values of $m_{\rm light}$. In addition, it turns out that $b$ and $c$ need to be small in order to keep the left-right mixing small enough,
since the rotation matrices in Eq.~\eqref{eq:lrmixing_md_diag} can lead to large entries in the $(1,1)$, $(1,2)$ and $(2,1)$ positions of $M^{}_DM_R^{-1}$, which enhance LFV processes.

\begin{figure}[tpb]
 \centering
 \includegraphics[angle=270,width=0.9\textwidth]{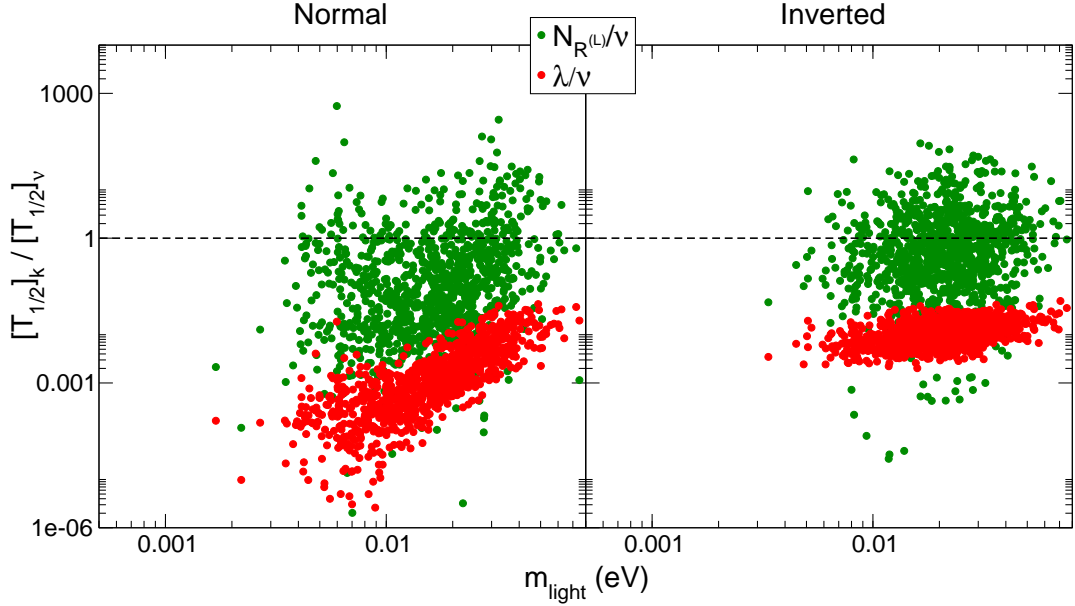}
 \caption{Ratio of half-life contributions, $[T^{0\nu}_{1/2}]_\lambda/[T^{0\nu}_{1/2}]_\nu$ and $[T^{0\nu}_{1/2}]_{N_R^L}/[T^{0\nu}_{1/2}]_\nu$.}
\label{fig:thalf_ratios}
\end{figure}

In order to ascertain whether one diagram might dominate over another it is interesting to look at the ratios of different halflives, which has the added advantage that uncertainties in NMEs will drop
out. In Fig.~\ref{fig:thalf_ratios} we show the ratios of various halflives to the standard half-life, calculated for the example texture. Here it is obvious that the $\lambda$-contribution can be
larger than the light neutrino contribution.

\section{Conclusion} \label{sec:concl}

In this paper we have investigated the interplay of neutrinoless double beta decay and charged lepton flavour violation in the context of the left-right symmetric model, paying particular attention to
those $\obb$ diagrams usually neglected in the literature. In the case of pure type~II seesaw we have shown that the triplet contribution to $\obb$ should not be neglected for all light neutrino
masses. For pure type I seesaw there exist regions of parameter space in which all diagrams can have similar orders of magnitude, which makes distinguishing the leading contribution difficult. In
particular, the momentum-dependent $\lambda$-diagram can be larger than expected. As we have shown, the bounds from lepton flavour violating decays complement the study of  lepton number violation,
and can be used to further restrict the parameter space. A comprehensive study should include the type I+II case, which we
leave for future work.

\csection{Acknowledgements}
~This work was supported by the Max Planck Society in the project MANITOP through the Strategic Innovation Fund. JB thanks Alexander Dueck, Julian Heeck and Tibor Frossard for useful discussions.

\csection{Appendix}
\appendix
\addcontentsline{toc}{chapter}{Appendix} 
\renewcommand{\theequation}{A-\arabic{equation}}
\setcounter{equation}{0}

\section{\texorpdfstring{Correlation between half-lives for $\obb$ in $^{76}$Ge and $^{136}$Xe}{Correlation between half-lives for 0nubb in 76Ge and 136Xe}} \label{sect:halflife_corrs}

\begin{figure}[htp]
 \centering
 \subfigure[light neutrino exchange]{\label{fig:halflife_corr_nu}
 \includegraphics[width=0.5\textwidth]{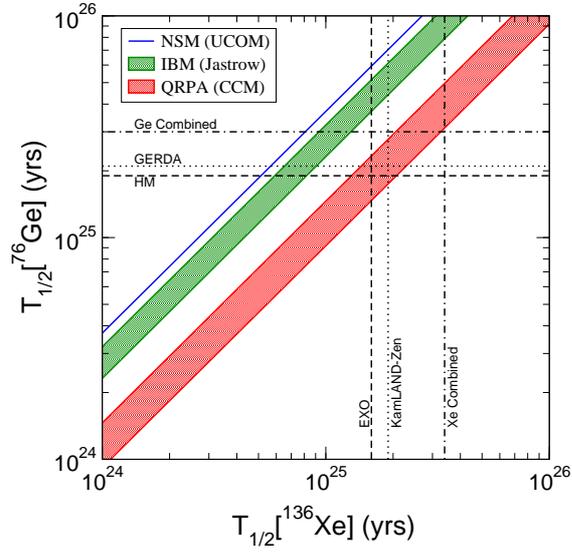}}\hspace{-1cm}
 \subfigure[heavy neutrino exchange]{\label{fig:halflife_corr_NR}
 \includegraphics[width=0.5\textwidth]{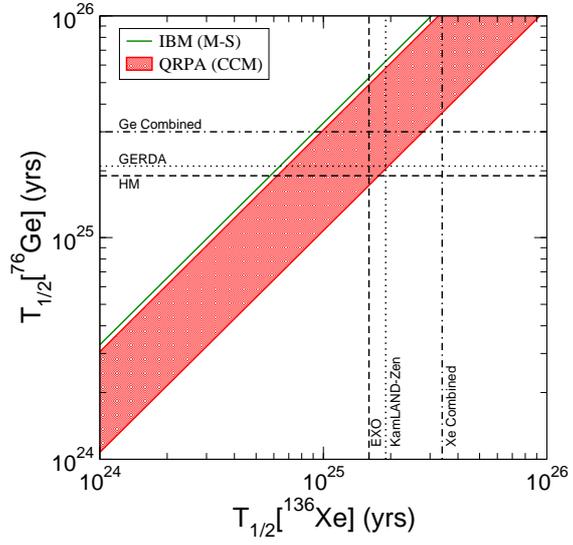}}
 \subfigure[$\lambda$-diagram]{\label{fig:halflife_corr_lam}
 \includegraphics[width=0.5\textwidth]{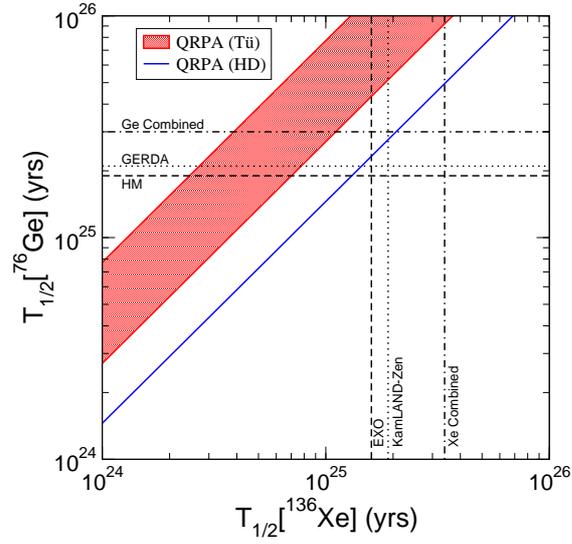}}\hspace{-1cm}
 \subfigure[$\eta$-diagram]{\label{fig:halflife_corr_eta}
 \includegraphics[width=0.5\textwidth]{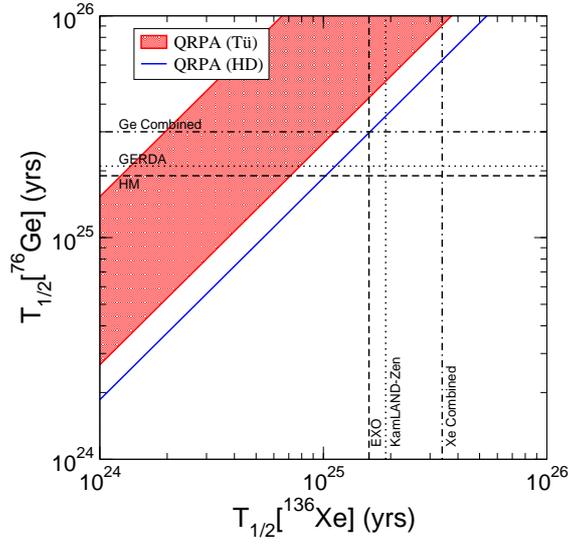}}
 \caption{Correlations between the $\obb$ half-lives in $^{76}$Ge and $^{136}$Xe for different matrix element calculations and particle physics contributions. The relevant limits from
Table~\ref{table:halflifelimits} are indicated by horizontal and vertical lines.}
 \label{fig:halflife_corrs}
\end{figure}

After the recent release of results from the GERDA experiment~\cite{Agostini:2013mzu} it is interesting to study the correlation between the $\obb$ half-lives in $^{76}$Ge and
$^{136}$Xe, for different matrix element calculations (see also Ref.~\cite{Gando:2012zm}). The current limits from the different experiments are given in Table~\ref{table:halflifelimits}. We have
plotted the correlations for light and heavy neutrino exchange as well as the $\lambda$- and $\eta$-diagrams in Fig.~\ref{fig:halflife_corrs}, using the matrix elements from
Tables~\ref{table:matrix_elements_2}, \ref{table:matrix_elements_3} and \ref{table:matrix_elements_4} together with the (new) phase space factor from the third column of
Table~\ref{table:matrix_elements}. The diagonal lines allow one to translate a half-life measured in $^{76}$Ge to one measured in $^{136}$Xe, and vice versa; the bands indicate the uncertainty in the
NMEs.

\begin{table}[htp]
 \centering
 \caption{Limits on the half-life of $\obb$ from different experiments.}
 \label{table:halflifelimits}
 \vspace{7pt}
 \begin{tabular}{lc}
  \hline \hline 
  \T Experiment & Limit $[10^{25}\ {\rm yrs}]$ \\[1mm]
  \hline HM & 1.9 \\
  GERDA & 2.1 \\
  Combined $^{76}$Ge & 3.0 \\
  \hline EXO & 1.6 \\
  KamLAND-Zen & 1.9 \\
  Combined $^{136}$Xe & 3.4 \\
  \hline \hline
 \end{tabular}
\end{table}

\begin{table}[htp]
 \centering
 \caption{$^{76}$Ge and $^{136}$Xe matrix elements for light neutrino exchange (${\cal M}^{0\nu}_\nu$) rescaled for $g_A = 1.25$ and $r_0 = 1.1$~fm.}
\label{table:matrix_elements_2}
\vspace{10pt}
 \begin{tabular}{lccc}
 \hline \hline
 \T Isotope & NSM (UCOM)~\cite{Menendez:2008jp} & QRPA (CCM)~\cite{Simkovic:2009pp} & IBM (Jastrow)~\cite{Barea:2009zza} \\
\hline \T
$^{76}$Ge & 2.58 &4.07--6.64 & 4.25--5.07 \\
$^{136}$Xe & 2.00 & 1.57--3.24 & 3.07 \\
\hline \hline
 \end{tabular}
\end{table}

\begin{table}[htp]
 \centering
 \caption{Same as Table~\ref{table:matrix_elements_2}, for heavy neutrino exchange (${\cal M}^{0\nu}_N$).}
\label{table:matrix_elements_3}
\vspace{10pt}
 \begin{tabular}{lcc}
 \hline \hline
 \T Isotope & IBM (M-S)~\cite{Barea:2013bz} & QRPA (CCM)~\cite{Faessler:2011rv} \\
\hline \T
$^{76}$Ge & 48.1 & 233--412 \\
$^{136}$Xe & 35.1 & 164--172 \\
\hline \hline
 \end{tabular}
\end{table}

\begin{table}[htp]
 \centering
 \caption{Same as Table~\ref{table:matrix_elements_2}, for the $\lambda$- and $\eta$-diagrams. The matrix elements ``QRPA (HD)'' were extracted from the limits given in
Ref.~\cite{Deppisch:2012nb}.}
\label{table:matrix_elements_4}
\vspace{10pt}
 \begin{tabular}{lcc|cc}
 \hline \hline
 \T \multirow{2}{*}{Isotope} & \multicolumn{2}{c|}{${\cal M}^{0\nu}_\lambda$} & \multicolumn{2}{c}{${\cal M}^{0\nu}_\eta$} \\
 & QRPA (CCM)~\cite{Pantis:1996py} & QRPA (HD)~\cite{Deppisch:2012nb} & QRPA (CCM)~\cite{Pantis:1996py} & QRPA (HD)~\cite{Deppisch:2012nb} \\[1mm]
\hline \T
$^{76}$Ge & 1.75--3.76 & 4.47 & 235--637 & 791\\
$^{136}$Xe & 1.96--2.49 & 2.17 & 370--419 & 434\\ 
\hline \hline
 \end{tabular}
\end{table}

\section{Details of lepton flavour violating expressions} \label{sect:lfv_lrsm}

Here we give details of the different contributions to lepton flavour violating processes. 

\subsection{Lagrangians \& couplings} \label{subsect:lr_lfv_exp}

LFV decays proceed via the charged current in Eq.~\eqref{eq:lag_cc_lr}, which we repeat here for convenience, as well as the couplings of the charged components of Higgs triplets to lepton doublets in Eq.~\eqref{eq:lag_full_lep}; the relevant terms are (with $h_L=h_R=h$)
\begin{align}
 {\cal L}^{\rm lep}_{CC} &=
\frac{g}{\sqrt{2}}\left[\overline{\ell'}\gamma^\mu P_L \nu'W_{L\mu}^- + \overline{\ell'}\gamma^\mu P_R \nu' W_{R\mu}^-\right] + {\rm h.c.}\,, \notag \\
 {\cal L}_{\delta^{\pm}_{L}} &= \frac{\delta_L^+}{\sqrt{2}}\left[\overline{{\nu'_L}^c}h\ell'_{L}+\overline{{\ell'_L}^c}h\nu'_L\right] +{\rm h.c.}\,,\\
 {\cal L}_{\delta^{\pm\pm}_{L,R}} &= \delta_{L,R}^{++}\overline{{\ell'}^c}h P_{L,R}\ell' + \delta_{L,R}^{--}\overline{\ell'}h^\dagger P_{R,L}{\ell}'^c\, . \notag 
\end{align}
Rotating the fields to the physical basis gives 
\begin{align}
 {\cal L}^{\rm lep}_{CC} &= \frac{g}{\sqrt{2}}\left[\overline{\ell_L}\gamma^\mu K_L n_L
(W_{1\mu}^- + \xi e^{i\alpha} W_{2\mu}^-) + \overline{\ell_R}\gamma^\mu  K_R n^c_L (-\xi e^{-i\alpha} W_{1\mu}^- + W_{2\mu}^-)\right]
+ {\rm h.c.}\,,  \notag \\
 {\cal L}_{H_1}& 
=\frac{g}{\sqrt{2}}\left[H_1^+\overline{n_L^c}\left(K_L^T\tilde{h}_L\right)\ell_L + H_1^-\overline{\ell_L}\left(\tilde{h}_L^\dagger K_L^*\right)n_L^c\right], \\
 {\cal L}_{\delta^{\pm\pm}_{L,R}} &= 
\frac{g}{2}\left[\delta^{++}_{L,R}\overline{\ell^c}\tilde{h}_{L,R}P_{L,R}\ell+\delta^{--}_{L,R}\overline{\ell}\tilde{h}_{L,R}^\dagger P_{R,L}\ell^c\right], \notag
\end{align}
where we have used Eqs.~\eqref{eq:md_ml_mr}, \eqref{eq:nu_rot_mass}, \eqref{eq:wlwr_mixing} and \eqref{eq:mw1_2}, with
\begin{equation}
 \tilde{h}_{L,R} \equiv (V^\ell_{L,R})^TV_R^\nu\frac{\tilde{M}_\nu}{\mwr}{V_R^\nu}^TV_{L,R}^\ell = (V^\ell_{L,R})^T\frac{M_R}{\mwr}V_{L,R}^\ell\,,
\end{equation}
and $\tilde{M}_\nu={\rm diag}(m_1,m_2,m_3,M_1,M_2,M_3)$. The LFV parameter is
\begin{equation}
 g^{L,R}_{\rm lfv} \equiv \left[\tilde{h}_{L,R}^\dagger \tilde{h}_{L,R}\right]_{e\mu} = \left[V_{L,R}^{\ell\dagger} {V_R^\nu}^*\left(\frac{\tilde{M}_\nu}{\mwr}\right)^2{V_R^\nu}^TV_{L,R}^\ell\right]_{e\mu} = \left[V_{L,R}^{\ell\dagger}\frac{M_R^*M_R}{\mwr^2}V_{L,R}^\ell\right]_{e\mu} .
\end{equation}

In the manifest left-right symmetry case (discrete parity symmetry), $V_L^\ell = V_R^\ell$, so that these expressions become \cite{Cirigliano:2004mv}
\begin{equation}
 \tilde{h} \equiv \tilde{h}_L=\tilde{h}_R = K_R^*\frac{\tilde{M}_\nu}{\mwr}K_R^\dagger\, , \quad {\rm and} \quad g_{\rm lfv} \equiv g^L_{\rm lfv} = g^R_{\rm lfv} = \left[K_R\left(\frac{\tilde{M}_\nu}{\mwr}\right)^2K_R^\dagger\right]_{e\mu}.
\label{eq:manifest_LRSM}
\end{equation}
In our case we take the charged lepton mixing matrices to be diagonal so that all processes depend on a combination of the mixing matrices $S$ and $V$ [see Eq.~\eqref{eq:lfv_params}], depending on the helicity of the different particles.

\subsection{Decay widths and branching ratios}

The effective Lagrangian for $\mu$ to $e$ conversion can be written as
\begin{equation}
\begin{split}
 {\cal L}_{\mue} = &-
\frac{e g^2}{4(4\pi)^2\mwl^2}m_\mu\overline{e}\sigma_{\mu\nu}(G^\gamma_LP_L+G^\gamma_RP_R)\mu F^{\mu\nu} \\ 
 &-\frac{\alpha_W^2}{2\mwl^2}\sum_q \left\{\overline{e}\gamma_\mu\left[W^q_LP_L+W^q_RP_R\right]\mu\; \overline{q}\gamma^\mu q\right\} + {\rm h.c.},
\end{split}
\end{equation}
with $\sigma_{\mu\nu} \equiv \frac{i}{2}[\gamma_\mu,\gamma_\nu]$ and the form factors $G^\gamma_{L,R}$ and $W^{u,d}_{L,R}$.
The full matrix element for $\muegam$ is given by
\begin{equation}
 \begin{split}
i{\cal M}(\muegam) &= \frac{e\alpha_W}{8\pi\mwl^2}\epsilon_\gamma^\mu \overline{e}\left[\left(q^2\gamma_\mu - q_\mu\slashed{q}\right)\left(F^\gamma_L P_L+F^\gamma_R P_R\right) \right. \\ & \left. -im_\mu\sigma_{\mu \nu}q^\nu\left(G^\gamma_L P_L+G^\gamma_R P_R\right)\right]\mu,
\end{split} 
\end{equation}
with the anapole and dipole form factors $F^{\gamma}_{L,R}$ and $G^\gamma_{L,R}$ defined in Eqs.~\eqref{eq:fgamm_form} and \eqref{eq:dipole_form_facts}.

The on-shell decay $\muegam$ only receives contributions from the $G^\gamma_{L,R}$ terms, the branching ratio turns out to be
\begin{equation}
  {\rm BR}_{\muegam} = \frac{\alpha_W^3s_W^2m_\mu^5}{256\pi^2\mwl^4\Gamma_\nu}\left(|G^\gamma_L|^2+|G^\gamma_R|^2\right) = \frac{3\alpha_{\rm em}}{2\pi}\left(|G^\gamma_L|^2+|G^\gamma_R|^2\right), 
\label{eq:brmuegam_RR_full}
\end{equation}
where
\begin{align}
\begin{split}
 \begin{split}
G^\gamma_{L} = \sum_{i=1}^3&\left\{V^{}_{\mu i}V^*_{ei}|\xi|^2G^{\gamma}_{1}(x_i) -S^*_{\mu i}V^*_{ei}\xi e^{-i \alpha} G^{\gamma}_2(x_i) \frac{M_i}{m_\mu} \right.\\
&\left.+ \ V^{}_{\mu i}V^*_{ei}\left[\frac{\mwl^2}{\mwr^2}G^{\gamma}_{1}(y_i)+\frac{2 y_i}{3}\frac{\mwl^2}{m_{\delta^{++}_R}^2}\right]\right\}, \end{split}\\[1mm]
 \begin{split}
G^\gamma_{R} = \sum_{i=1}^3&\left\{S^*_{\mu i} S^{}_{ei} G^{\gamma}_{1}(x_i) - V_{\mu i} S_{ei}\xi e^{i \alpha}G^{\gamma}_ 2(x_i)\frac{M_i}{m_\mu} \right. \\
&\left.+ \ V^{}_{\mu i} V^*_{ei}\,y_i\left[\frac{2}{3}\frac{\mwl^2}{m_{\delta^{++}_L}^2}+\frac{1}{12}\frac{\mwl^2}{m^2_{H_1^+}}\right]\right\}, \end{split} \end{split} \label{eq:dipole_form_facts}
\end{align}
with $x_i \equiv (M_i/\mwl)^2$, $y_i \equiv (M_i/\mwr)^2$ and the loop functions $G^\gamma_{1,2}(x)$ defined in Eq.~\eqref{eq:loop_funcs}. In addition, the electric dipole moment of charged lepton $\ell_\alpha$ ($\alpha = e,\mu,\tau$) is given by \cite{Nieves:1986uk,Nemevsek:2012iq,Tello_thesis:2012}
\begin{equation}
 d_\alpha = \frac{e\,\alpha_W}{8\pi\mwl^2}{\rm Im}\left[\sum_{i=1}^3 S_{\alpha i} V_{\alpha i}\xi e^{i \alpha}G^{\gamma}_ 2(x_i)M_i\right],
\label{eq:edipole}
\end{equation}
which is similar to the mixed diagram contribution in $\muegam$.

The tree level contribution to $\mueee$ in Eq.~\eqref{eq:brmueee_tree} can be rewritten as 
\begin{align}
  {\rm BR}^{\rm triplet}_{\mueee} &= \frac{\alpha_W^4m_\mu^5}{24576\pi^3\mwl^4\Gamma_\mu}\frac{(4\pi)^2}{2\alpha_W^2}\left|\tilde{h}^{}_{\mu e}\tilde{h}^*_{ee}\right|^2\left(\frac{\mwl^4}{m_{\delta^{++}_L}^4}+\frac{\mwl^4}{m_{\delta^{++}_R}^4} \right),
\end{align}
to be compared with the loop-suppressed type I seesaw contribution given by~\cite{Ilakovac:1994kj,Ilakovac:2012sh}
\begin{equation}
 \begin{split}
 {\rm BR}^{\rm type\ I}_{\mueee} &= \frac{\alpha_W^4 m_\mu^5}{24576\pi^3\mwl^4\Gamma_\mu} \left\{ 2\left[\left| \frac{1}{2} B^{\mu eee}_{LL} + F^{Z_1}_L - 2s_W^2(F^{Z_1}_L-F^\gamma_L) \right|^2 + \left| \frac{1}{2} B^{\mu eee}_{RR} - 2s_W^2(F^{Z_1}_R-F^\gamma_R) \right|^2 \right]\right. \\[1mm]
& \left. +\left|2s_W^2(F^{Z_1}_L - F^\gamma_L)-B^{\mu eee}_{LR}\right|^2 +\left|2s_W^2(F^{Z_1}_R - F^\gamma_R)-(F^{Z_1}_R+B^{\mu eee}_{RL})\right|^2 \right. \\[1mm] & \left. +8s_W^2\left[{\rm Re}\left((2F^{Z_1}_L+B^{\mu eee}_{LL}+B^{\mu eee}_{LR}){G^\gamma_R}^*\right) + {\rm Re}\left((F^{Z_1}_R+B^{\mu eee}_{RR}+B^{\mu eee}_{RL}){G^\gamma_L}^*\right)\right] \right. \\[1mm] & \left. -48s_W^4\left[{\rm Re}\left((F^{Z_1}_L-F^\gamma_L){G^\gamma_R}^*\right)+{\rm Re}\left((F^{Z_1}_R-F^\gamma_R){G^\gamma_L}^*\right)\right] \right. \\[1mm] & \left. + 32s_W^4\left(\left|G^\gamma_L\right|^2+\left| G^\gamma_R\right|^2\right)\left[\ln{\frac{m_\mu^2}{m_e^2}}-\frac{11}{4}\right]\right\}. \label{eq:brmueee_LL_full} \end{split} 
\end{equation}
The interference terms between triplet exchange and gauge boson mediated loop and box diagrams are
\begin{align}
\begin{split}
 {\rm BR}^{\rm triplet+type\ I}_{\mueee} &= \frac{\alpha_W^4 m_\mu^5}{24576\pi^3\mwl^4\Gamma_\mu} \frac{2(4\pi)}{\alpha_W} \ \times \\ & \left\{\frac{\mwl^2}{\mdl^2} {\rm Re}\left[ 2s_W^2 T^*F^\gamma_L+4s_W^2 T^*G^\gamma_R+T^*B^{\mu eee}_{LL}+T^*F_L^{Z_1}(1-2s_W^2)\right] \right. \\
 &\left. + \frac{\mwl^2}{\mdr^2} {\rm Re}\left[ 2s_W^2 T^*F^\gamma_R+4s_W^2 T^*G^\gamma_L+T^*B^{\mu eee}_{RR}-2s_W^2T^*F_R^{Z_1}\right]\right\}, \end{split}
\end{align}
where $T \equiv \tilde{h}_{\mu e}\tilde{h}^*_{ee}$ and $\tilde{h}_{\alpha \beta}$ is defined in Eq.~\eqref{eq:lfv_params}. Note that the triplet term effectively has the same structure as the box contribution (after Fierz transformations, see Ref.~\cite{Cuypers:1996ia}), so we expect it to interfere with the other amplitudes in the same way.

The form factors for off-shell photon exchange are
\begin{align}
\begin{split}
 F_L^\gamma &= \sum_{i=1}^3\left\{ S^*_{\mu i}S_{e i}F_{\gamma}(x_i) - 
V_{\mu i}V^*_{ei}\,y_i\left[\frac{2}{3}\frac{\mwl^2}{m_{\delta^{++}_L}^2}\ln\frac{m_\mu^2}{m_{\delta^{++}_L}^2}+\frac{1}{18}\frac{\mwl^2}{m^2_{H_1^+}}\right]\right\}, \\[1mm] 
F_R^\gamma &= \sum_{i=1}^3 V_{\mu i}V^*_{ei} \left[|\xi|^2F_\gamma(x_i)+\frac{\mwl^2}{\mwr^2}F_\gamma(y_i) - y_i\frac{2}{3}\frac{\mwl^2}{\mdr^2}\ln\frac{m_\mu^2}{\mdr^2}\right], \label{eq:fgamm_form}
\end{split} 
\end{align}
where the logarithmic term is a simplified version of the usual triplet loop function \cite{Raidal:1997hq}, since we take the doubly charged scalar mass to be much larger than the charged lepton masses ($m_{\delta_{L,R}} \gg m_{e,\mu,\tau}$). The $Z_1$-boson exchange terms\footnote{We ignore terms from the exchange of the heavier $Z_2$ boson.} can be expressed as
\begin{align}
\begin{split}
 F_L^{Z_1} &= \sum_{i,j=1}^3 S^*_{\mu i}S^{}_{ej}\left\{\delta_{ij}\left(F_Z(x_i)+2G_Z(0,x_i)\right) \right. \\
 & \left. + \ (S^TS^*)_{ij}\left[G_Z(x_i,x_j)-G_Z(0,x_i)-G_Z(0,x_j)\right] + (S^\dagger S)_{ij}H_Z(x_i,x_j)\right\}, \\
 F_R^{Z_1} &\simeq \sum_{i=1}^3 V_{\mu i}V^*_{ei} \left[\frac{1-2s_W^2}{2c_W^2}\frac{\mwl^2}{\mwr^2}\left(F_Z(y_i)+2G_Z(0,y_i)-\frac{y_i}{2}\right) \right. \\ & \left. +\frac{\mwl^2}{\mwr^2}D_Z(y_i,x_i)+\frac{\mwl^2}{\mwr^2}D_Z(y_i,z_i)\right]. \end{split} \label{eq:z_form_fact}
\end{align}
where $z_i=(M_i/m_{H_2})^2$; the box diagram form factors are\footnote{We neglect terms proportional to $|\xi|^2$.}
\begin{align}
\begin{split}
 B^{\mu eee}_{LL} &= -2\sum_{i=1}^3\left\{ S^*_{\mu i}S^{}_{ei} \left[F_{\rm X box}(0,x_i) - F_{\rm X box}(0,0)\right] \right\} \\ 
 & + \sum_{i,j=1}^3S^*_{\mu i}S^{}_{e j}\left\{-2S^*_{e j}S^{}_{e i}\left[F_{\rm X box}(x_i,x_j)-F_{\rm X box}(0,x_j)-F_{\rm X box}(0,x_i)+F_{\rm X box}(0,0)\right] \right. \\
 & \left. + \ S^*_{e i}S^{}_{e j} G_{\rm box}(x_i,x_j,1)\right\},
\end{split} \\[1mm]
\begin{split}
 B^{\mu eee}_{RR} &= -2\frac{\mwl^2}{\mwr^2}\sum_{i=1}^3\left\{ V^{}_{\mu i}V^*_{ei}  \left[F_{\rm X box}(0,y_i) - F_{\rm X box}(0,0)\right] \right\} \\ 
 & + \sum_{i,j=1}^3V^{}_{\mu i}V^{*}_{e j}\left\{-2V^{}_{e j}V^{*}_{e i}\left[F_{\rm X box}(y_i,y_j)-F_{\rm X box}(0,y_j)-F_{\rm X box}(0,y_i)+F_{\rm X box}(0,0)\right] \right. \\
 & \left. + \ V^{}_{e i}V^{*}_{e j} G_{\rm box}(y_i,y_j,1)\right\},
\end{split} \label{eq:box_form_facts_mueee}
\end{align}
for purely left- and right-handed contributions and
\begin{align}
 B^{\mu eee}_{LR} &= \frac{1}{2}\frac{\mwl^2}{\mwr^2}\sum_{i,j=1}^3 S^*_{\mu i} S^{}_{e j} V^{}_{e i} V^{*}_{e j}  G_{\rm box}\left(x_i,x_j,\frac{\mwl^2}{\mwr^2}\right), \\[1mm]
B^{\mu eee}_{RL} &= \frac{1}{2}\frac{\mwl^2}{\mwr^2}\sum_{i,j=1}^3 V^{}_{\mu i} V^{*}_{e j} S^{*}_{e i} S^{}_{e j}  G_{\rm box}\left(x_i,x_j,\frac{\mwl^2}{\mwr^2}\right),
\end{align}
for diagrams with mixed helicity. The loop-suppressed amplitudes with right-handed currents contain the ${\cal O}(1)$ mixing matrix $V$ as well as the additional suppression factor of $(\mwl/\mwr)^2$; without the enhancement from large left-right mixing (in $S$), we expect those contributions to be much smaller than the tree level one in Eq.~\eqref{eq:brmueee_tree}. The mixed left-right box contributions come from an effective four fermion operator, as is the case in kaon mixing \cite{Mohapatra:1983ae,Zhang:2007da,Maiezza:2010ic}, with a factor of $1/2$ coming from the Fierz transformation of a scalar to vector contribution (see Ref.~\cite{Ilakovac:2012sh}).

$\mue$ conversion in nuclei is similar to $\mueee$ and receives contributions from the same loop and box diagrams.\footnote{Although the process can also be mediated at tree-level by neutral Higgs bosons, these particles have to be very heavy due to constraints from $K^0$-$\overline{K}^0$ mixing.} The $\mue$ conversion rate is given by \cite{Cirigliano:2004mv,Pilaftsis:2005rv,Alonso:2012ji,Ilakovac:2012sh}
\begin{equation}
 {\rm R}^{A(N,Z)}_{\mue} = \frac{\alpha_{\rm em}^3\alpha_W^4m_\mu^5}{16\pi^2\mwl^4\Gamma_{\rm capt}}\frac{Z_{\rm eff}^4}{Z}\left|F(-m_\mu^2)\right|^2 \left(|Q_L^W|^2+|Q_R^W|^2\right), \label{eq:brmue_full}
\end{equation}
where
\begin{equation}
 Q^W_{L,R} = (2Z+N)\left[W^u_{L,R}-\frac{2}{3}s_W^2 G^\gamma_{R,L}\right] + (Z+2N)\left[W^d_{L,R} +\frac{1}{3}s_W^2 G^\gamma_{R,L} \right],
\end{equation}
and
\begin{align}
\begin{split}
W^u_{L,R} &= \frac{2}{3}s_W^2F^\gamma_{L,R}+\left(-\frac{1}{4}+\frac{2}{3}s_W^2\right)F^{Z_1}_{L,R}+\frac{1}{4}\left(B^{\mu e uu}_{LL,RR}+B^{\mu e uu}_{LR,RL}\right),\\ 
W^d_{L,R} &= -\frac{1}{3}s^2_WF^\gamma_{L,R}+\left(\frac{1}{4}-\frac{1}{3}s_W^2\right)F^{Z_1}_{L,R}+\frac{1}{4}\left(B^{\mu e dd}_{LL,RR}+B^{\mu e dd}_{LR,RL}\right),
\end{split}
\end{align}
are composite form factors. Note that the expression in Eq.~\eqref{eq:brmue_full} is derived by approximating all interactions to be point-like and taking the proton and neutron densities to be equal. In this case the wavefunction overlap integrals $D$ and $V^{(p,n)}$ calculated in Ref.~\cite{Kitano:2002mt} can be replaced by the quantities $Z_{eff}$ and the form factor $F(-m_\mu^2)$, where
\begin{equation}
 \frac{V^{(p)}}{\sqrt{Z}} = \frac{Z_{eff}^2F(-m_\mu^2)\alpha_{\rm em}^{\frac{3}{2}}}{4\pi}\, ,
\end{equation}
and $V^{(p)}/Z \simeq V^{(n)}/N$. The relevant box diagram form factors are
\begin{align}
\begin{split}
B^{\mu e uu}_{LL} &= \sum_{i=1}^3S^*_{\mu i}S^{}_{e i}\left[F_{\rm box}(0,x_i) - F_{\rm box}(0,0)\right], \\
B^{\mu e dd}_{LL} &\simeq \sum_{i=1}^3S^*_{\mu i}S^{}_{e i}\left\{F_{\rm X box}(0,x_i) - F_{\rm X box}(0,0)  \right. \\ 
& \left. + |V_{td}|^2\left[F_{\rm X box}(x_t,x_i)-F_{\rm X box}(0,x_i)-F_{\rm X box}(0,x_t)+F_{\rm X box}(0,0)\right]\right\}, \\
B^{\mu e qq}_{RR} &= \frac{\mwl^2}{\mwr^2}B^{\mu e qq}_{LL}(S^{} \leftrightarrow V^*\,;\,x_i \leftrightarrow y_i\,;\,x_t \leftrightarrow y_t)\, ,
\end{split}
\end{align}
where $x_t = m_t^2/\mwl^2$ and $y_t = m_t^2/\mwr^2$.

Finally we note that the presence of non-unitary mixing in the light neutrino sector (due to the matrix $S \simeq M^{}_DM_R^{-1}$) also affects the standard muon decay width, $\Gamma_\mu$ (and thus the determination of $G_F$), as well as the capture rate for muons on the nucleus, $\Gamma_{\rm capt}$. Explicitly, one has
\begin{equation}
 \Gamma_\mu \simeq \Gamma_\mu^{(0)}\left(\mathbb{1} - [S S^\dagger]_{ee} - [SS^\dagger]_{\mu\mu}\right) \quad {\rm and} \quad
 \Gamma_{\rm capt} \simeq \Gamma_{\rm capt}^{(0)}\left(\mathbb{1} - [SS^\dagger]_{\mu\mu}\right),
\end{equation}
where $\Gamma_\mu^{(0)}$ and $\Gamma_{\rm capt}^{(0)}$ are the SM values and we have omitted terms of order $S^4$. These expressions occur in the denominators of the branching ratio formulae in Eq.~\eqref{eq:br_defs}, and since the numerators are in general proportional to ${\cal O}(S^4)$ the effect will be negligible; in our analysis we use the standard value $\Gamma_\mu = G_F^2m_\mu^5/(192\pi^3)$.

\subsection{Loop functions} \label{sect:form_loop}

The relevant loop functions are
\begin{align} \label{eq:loop_funcs}
 \begin{split}
 F_{\gamma}(x) &= \frac{7x^3-x^2-12x}{12(1-x)^3}-\frac{x^4-10x^3+12x^2}{6(1-x)^4}\ln{x}, \\[1mm]
 G^{\gamma}_1(x) &= -\frac{2x^3+5x^2-x}{4(1-x)^3} - \frac{3x^3}{2(1-x)^4}\ln{x}, \\[1mm]
 G^{\gamma}_2(x) &= \frac{x^2-11 x+4}{2 (1-x)^2}-\frac{3 x^2}{(1-x)^3}\ln{x}, \\[1mm]
 F_Z(x) &= -\frac{5x}{2(1-x)}-\frac{5x^2}{2(1-x)^2}\ln{x}\,, \\[1mm] 
 G_Z(x,y) &= -\frac{1}{2(x-y)}\left[\frac{x^2 (1-y)}{1-x}\ln{x}- \frac{y^2(1-x)}{1-y}\ln{y}\right], \\[1mm]
H_Z(x,y) &= \frac{\sqrt{x y}}{4(x-y)} \left[\frac{x^2-4 x}{1-x}\ln{x} - \frac{y^2-4y}{1-y}\ln{y}\right], \\[1mm]
 D_Z(x,y) &= x \left(2-\ln\frac{y}{x}\right)+\frac{(-8 x+9 x^2-x^3)+(-8 x^2+x^3) \ln x}{(1-x)^2}+\frac{x(y-y^2+y^2 \ln y)}{(1-y)^2} \\ &+\frac{2 x y(4-x) \ln x}{(1-x) (1-y)}+\frac{2 x (x-4 y) \ln \frac{y}{x}}{(1-y) (x-y)}\, , \\[1mm]
 F_{\rm box} &= \left(4+\frac{xy}{4}\right)I_2(x,y,1)-2xyI_1(x,y,1), \\[1mm]
 F_{\rm X box}(x,y) &= -\left(1+\frac{xy}{4}\right)I_2(x,y,1)-2xyI_1(x,y,1), \\[1mm] 
 G_{\rm box}(x,y,\eta) &= -\sqrt{x y} \left[(4+x y \eta) I_1(x,y,\eta)-(1+\eta)I_2(x,y,\eta)\right], \\
 \end{split} 
\end{align}
where
\begin{align}
\begin{split}
 I_1(x,y,\eta) &= \left[\frac{x \ln{x}}{(1-x) (1-\eta  x) (x-y)}+(x \leftrightarrow y)\right] -\frac{\eta  \ln{\eta} }{(1-\eta ) (1-\eta  x)   (1-\eta  y)}\, , \\
I_2(x,y,\eta) &= \left[\frac{x^2 \ln{x}}{(1-x) (1-\eta  x) (x-y)}+(x \leftrightarrow y)\right] -\frac{  \ln{\eta} }{(1-\eta ) (1-\eta  x)   (1-\eta  y)}\, , \\
I_i(x,y,1) &\equiv \lim_{\eta\to 1}I_{i}(x,y,\eta)\,,
\end{split}
\end{align}
and the limiting values are
\begin{align}
\begin{split}
G_Z(0,x) &= -\frac{x\ln{x}}{2(1-x)}\,, \\[1mm]
F_{\rm box}(0,x) &= \frac{4}{1-x}+\frac{4x}{(1-x)^2}\ln{x}\, , \\[1mm]
F_{\rm X box}(0,x) &= -\frac{1}{1-x}-\frac{x\ln{x}}{(1-x)^2}\, . \end{split}
\end{align}

\begin{landscape}
\section{Explicit numerical example} \label{sect:explicit_example}

Here we give an explicit numerical example for the case of type I dominance and normal neutrino mass ordering, following the ansatz of Ref.~\cite{Mitra:2011qr} and fulfilling the bounds from LFV
experiments (see Section~\ref{subsect:lfv_lrsm}). All dimensionful parameters are given
in eV, unless otherwise indicated. From Eq.~\eqref{eq:texture_example}, the parameters
\begin{footnotesize}
\begin{align}
\begin{split}
 a &= 9.53381960582404819 
 \times 10^{-2} +  0.11713054331122945 
 i, \ b = 0.21843620328064534 
 - 0.22040775144734739 
 i,  \\
 c &= 4.31908935642526456 
 \times 10^{-2} -  2.92388739170286211 
 \times 10^{-4} i, \ d = -9.05724681558278330 
 \times 10^{-3} + 0.12019026634023072 
 i,  \\
 e &= 0.28217599917126424 
 - 0.17450535348840202 
 i, \ f = 0.84027958230331323 
 + 0.40461526271813769 
 i, \\
 g &= 0.30944075406011140 
 + 0.29546133055037482 
 i, \ h = 0.33205857612068856 
 + 0.82492777937530726 
 i,  \\ 
 j &= 0.63203076828617810 
 - 0.72194951521080608 
 i, \ \epsilon = -4.25119854705781844 
 \times 10^{-6} - 2.87020754827289911 
 \times 10^{-6} i,  \\
\kappa_+ &= 173.99999692800000 
\ {\rm GeV}, \ M = 676.84091139837646 
\ {\rm GeV}, 
\end{split}
\end{align}
\end{footnotesize}
lead to the matrices
\begin{scriptsize} 
\begin{align}
 \tilde{M}_D &= {\rm diag}(-0.334219074474+0.605264418163 i,-271654.304377+53946.936992 i,7.51521534750\times 10^9-5.087563972\times 10^7 i) \ {\rm eV}, \\ \quad {M'_R}^{-1} &=
\begin{pmatrix} -1.33816479812\times 10^{-14}+1.77575356803\times 10^{-13}  i & 4.16901511743\times 10^{-13}-2.57823294292\times 10^{-13}  i & 1.24147280129\times
   10^{-12}+ 5.97799654105\times 10^{-13}  i \\
 4.16901511743\times 10^{-13}- 2.57823294292\times 10^{-13}  i & 4.57183880065\times 10^{-13}+ 4.36529951979\times 10^{-13}  i & 1.41254316443\times
   10^{-18}- 6.58944920978\times 10^{-18}  i \\
 1.24147280129\times 10^{-12}+ 5.97799654105\times 10^{-13}  i & 1.41254316443\times 10^{-18}- 6.58944920978\times 10^{-18}  i & 3.52135444737\times
   10^{-23}+ 1.22979726454\times 10^{-23}  i \end{pmatrix} \ {\rm eV}^{-1},
\end{align}
\end{scriptsize}
which give the neutrino mass matrix 
\begin{footnotesize}
\begin{equation}
 m_\nu' = \begin{pmatrix} -7.52513003231\times 10^{-14}+ 3.98042836036\times 10^{-14}  i & 2.28019692342\times 10^{-8}+ 9.10546573825\times 10^{-8}  i &
+0.00580938497446-0.00418508532318 i \\
 +2.28019692342\times 10^{-8}+ 9.10546573825\times 10^{-8}  i & -0.0452024610799-0.0175437858823 i & 0.000117304047425-0.0140267422225 i \\
 0.00580938497446-0.00418508532318 i & 0.000117304047425-0.0140267422225 i & -0.00199811972713-0.000667611555085 i \end{pmatrix} \ {\rm eV}
\end{equation}
\end{footnotesize}
via Eq.~\eqref{eq:mnu_mddiag}. After diagonalizing $m_\nu'$ and rotating by $V_\nu$ [see Eq.~\eqref{eq:mnuflav}], the neutrino mass matrix in the flavour basis is
\begin{footnotesize}
\begin{equation}
 m_\nu = \begin{pmatrix} 0.00140944908669+0.00384187592338 i & -0.00347531018948+0.00895104270924 i & 0.00385457629281+0.00345336242992 i \\
 -0.00347531018948+0.00895104270924 i & -0.00226255970351+0.0301706233567 i & 0.000129801864554+0.0224710709065 i \\
 0.00385457629281+0.00345336242992 i & 0.000129801864554+0.0224710709065 i & -0.00433575002153+0.0255482257202 i \end{pmatrix} \ {\rm eV},
\end{equation}
\end{footnotesize}
with the eigenvalues
\begin{equation}
  m_1 = 0.00467695990924 \ {\rm eV}, \quad m_2 = 0.010179233482 \ {\rm eV}, \quad m_3 = 0.0522115758358 \ {\rm eV},
\end{equation}
The modified Dirac mass matrix is
\begin{footnotesize}
\begin{align}
 M_D &= \begin{pmatrix} -1.32139207855\times 10^8+ 3.41289155506\times 10^7  i & -4.59076639803\times 10^7- 7.24799214279\times 10^8  i & 4.00927912892\times
10^8+ 5.64722783204\times 10^8  i \\  -4.59076639803\times 10^7- 7.24799214279\times 10^8  i & 3.8343013446\times 10^9+ 4.86201875611\times 10^8  i & -3.35280821262\times 10^9+ 1.52965177159\times
10^9  i \\  4.00927912892\times 10^8+ 5.64722783204\times 10^8  i & -3.35280821262\times 10^9+ 1.52965177159\times 10^9  i & 1.95139172515\times 10^9- 2.92228131465\times 10^9  i
\end{pmatrix} \ {\rm eV},
\end{align}
\end{footnotesize}
and the final right-handed neutrino mass matrix is
\begin{footnotesize}
 \begin{equation}
  M_R = \begin{pmatrix}  -6.89802539588\times 10^{10}+ 1.95517241581\times 10^{11}  i & -8.25178556559\times 10^{11}- 1.6454840933\times 10^{11}  i & 2.17944743926\times
10^{11}- 2.70953080859\times 10^{11}  i \\  -8.25178556559\times 10^{11}- 1.6454840933\times 10^{11}  i & -7.72367515685\times 10^{11}- 8.54394798711\times 10^{11}  i & -5.59927219536\times
10^{11}- 4.48664481584\times 10^{11}  i    \\  2.17944743926\times 10^{11}- 2.70953080859\times 10^{11}  i & -5.59927219536\times 10^{11}- 4.48664481584\times 10^{11}  i & -4.59009131535\times
10^{11}+ 7.48335978813\times 10^{10}  i \end{pmatrix} \ {\rm eV},
 \end{equation}
\end{footnotesize}
with the eigenvalues 
\begin{equation}
 M_1 = 651.474530033 \ {\rm GeV}, \quad M_2 = 697.492992124 \ {\rm GeV}, \quad M_3 = 1833.67403677 \ {\rm GeV}.
\end{equation}

The left-right mixing is given by
\begin{footnotesize}
 \begin{equation}
  M^{}_DM_R^{-1} = \begin{pmatrix} 0.000149644130868+0.00133067594536 i & -0.000237728149067-0.000173576087253 i & 0.000139285740058-0.000220000171375 i \\
 -0.00710439002177-0.000550859287664 i & 0.0011451580917-0.00106845293524 i & 0.00101008111999+0.000948346776121 i \\
 0.00603697434636-0.00311752861487 i & -0.000392690432293+0.00144086921243 i & -0.00129494996783-0.000262316931666 i \end{pmatrix},
 \end{equation}
\end{footnotesize}
which is also one of the solutions of the equation $M^{}_DM_R^{-1} = i \sqrt{m^{}_\nu M_R^{-1}}$.
\end{landscape}

\addcontentsline{toc}{chapter}{References} 
\bibliographystyle{utcaps_mod}
\bibliography{lrbib.bib}

\end{document}